\documentclass[pdflatex,sn-mathphys-num]{sn-jnl}

\usepackage{graphicx}%
\usepackage{multirow}%
\usepackage{amsmath,amssymb,amsfonts}%
\usepackage{amsthm}%
\usepackage{mathrsfs}%
\usepackage[title]{appendix}%
\usepackage{xcolor}%
\usepackage{textcomp}%
\usepackage{manyfoot}%
\usepackage{booktabs}%
\usepackage{algorithm}%
\usepackage{algorithmicx}%
\usepackage{algpseudocode}%
\usepackage{listings}%

\usepackage{float}
\usepackage[T1]{fontenc}
\usepackage{textgreek}
\usepackage{hyperref}
\usepackage{multicol}

\newcommand{\MACS}{\text{MACS}_{30}}

\theoremstyle{thmstyleone}%
\theoremstyle{thmstyletwo}%

\theoremstyle{thmstylethree}%

\begin{document}

\title[New measurement of $^{50}$Cr and $^{53}$Cr (n,$\gamma$) cross sections at n\_TOF: a call for chromium nuclear data revision]{New measurement of $^{50}$Cr and $^{53}$Cr (n,$\gamma$) cross sections at n\_TOF: a call for chromium nuclear data revision}








\author[1,2]{P.~P\'{e}rez-Maroto} %
\author*[1,2]{C.~Guerrero}\email{cguerrero4@us.es} %
\author[3]{A.~Casanovas} %
\author[1,2]{B.~Fern\'{a}ndez} %
\author[4]{E.~Mendoza} %
\author[4]{V.~Alcayne} %
\author[5]{J.~Lerendegui-Marco} %
\author[6]{C.~Domingo-Pardo} %
\author[1]{ J.~M.~Quesada } %
\author[6]{ R.~Capote } %
% ...
%

\affil[1]{Universidad de Sevilla, Spain} %
\affil[2]{Centro Nacional de Aceleradores (CNA), Spain} %
\affil[3]{Universitat Polit\`{e}cnica de Catalunya, Spain} %
\affil[4]{Centro de Investigaciones Energ\'{e}ticas Medioambientales y Tecnol\'{o}gicas (CIEMAT), Spain} %
\affil[5]{Instituto de F\'{\i}sica Corpuscular, CSIC - Universidad de Valencia, Spain} %
\affil[6]{NAPC–Nuclear Data Section, International Atomic Energy Agency, Austria} %
% ...
%


\abstract{
$^{50}$Cr and $^{53}$Cr are very relevant in criticality safety benchmarks related to nuclear reactors. The discrepancies of up to 30\% between the neutron capture cross section evaluations have an important effect on the $k_{eff}$ and $k_{\infty}$ in criticality benchmarks particularly sensitive to chromium. In this work, the $^{50,53}$Cr(n,$\gamma$) cross sections are to be determined between 1 and 100~keV with an 8-10\% accuracy following the requirements of the NEA High Priority Request List (HPRL) to solve the current discrepancies. We have measured these reactions by the time-of-flight technique at the EAR1 experimental area of the n\_TOF facility, using an array of four C$_6$D$_6$ detectors with very low neutron sensitivity. The highly-enriched samples used are significantly thinner than in previous measurements, thus minimizing the multiple-scattering effects. We have produced, and analysed with the R-matrix analysis code SAMMY, capture yields featuring 33 resonances of $^{50}$Cr and 51 of $^{53}$Cr with an accuracy between 5\% and 9\%, hence fulfilling the requirements made by the NEA. The differential and integral cross sections have been compared to previous data and evaluations. The new measured $^{50,53}$Cr(n,$\gamma$) cross sections provide a valuable input for upcoming evaluations, which are deemed necessary given that the results presented herein do not support the increase in both cross sections proposed in the recent INDEN evaluation.
}

\maketitle

\section{Introduction}
\label{sect_intro}
In a context of increasing energy demand worldwide, nuclear energy has been recently recognized by the European Commission crucial to reduce greenhouse gases emissions~\cite{european2022complementary}. This framework requires nuclear energy to progress in terms of efficiency and safety performance, which needs to be addressed through a large effort on \textit{criticality safety} research programs~\cite{bess2020current}. 
 
In nuclear reactors, chromium is an important component (11-26\% abundance) of the stainless steel used as a structural material, and thus its neutron-interaction cross section plays an important role in the associated neutronic calculations. In particular, Trkov~\cite{trkov2018benchmarking} reported significant discrepancies between calculations and integral measurements when performing $k_{eff}$ and $k_\infty$ criticality benchmarks sensitive to chromium like HEU-COMP-INTER-005/4=KBR-15/Cr or PU-MET-INTER-002. These are due to discrepancies of about 30\% between the chromium cross sections libraries/evaluations found at the time. These findings led to a new entry in the Nuclear Energy Agency (NEA) High Priority Request List (HPRL)~\cite{dupont2020hprl} calling for a new neutron capture cross section measurement of $^{50}$Cr and $^{53}$Cr between 1 and 100~keV with an accuracy of 8-10\%.

The need of new measurements is related to the limited accuracy and experimental issues of the previous ones. Chromium is a relatively light isotope, and it is characterized by its large scattering-to-capture cross section ratio, especially at neutron energies $E_n$ between 1 and 10~keV, where several resonances with a large scattering cross section are found. They are sometimes referred to as a \textit{s}-wave resonance \textit{cluster}. Therefore, the neutron capture and transmission measurements of $^{50,53}$Cr made with relatively thick samples in the last decades of the XX century ~\cite{stieglitz1971kev,beer1975kev,kenny1977neutron,brusegan1986high} suffered from high sensitivity to scattered neutrons and strong multiple-scattering effects. More recently, in 2011 a neutron capture and transmission measurement of natural and $^{53}$Cr was performed by Guber et al.~\cite{guber2011neutron} at ORNL ORELA using low neutron sensitivity detectors, but the multiple-scattering was still an issue, as discussed later by Nobre et al.~\cite{nobre2021newly}. Considering the capture and transmission data from Ref.~\cite{guber2011neutron} together with another (unpublished) ORELA transmission data from Harvey et al., Leal et al.~\cite{leal2011evaluation} produced a cross section evaluation that has been adopted in JEFF-3.3~\cite{plompen2020joint}, ENDF/B-VIII.0~\cite{Brown20181}, JENDL-5~\cite{iwamoto2023japanese} and BROND-3.1~\cite{blokhin2016new}; while CENDL-3.2~\cite{ge2020cendl} is based on the data published before 2011. Only a few years ago the new INDEN evaluation~\cite{indenweb} has been released (and adopted in ENDF/B-VIII.1~\cite{nobre2024progress}), proposing a significant increase of the capture cross section of both $^{50}$Cr and $^{53}$Cr based on a re-analysis by Nobre et al.~\cite{nobre2021newly} of the data from Stieglitz et al.~\cite{stieglitz1971kev} and Guber et al.~\cite{guber2011neutron} in which issues with the sample thickness are identified and addressed through detailed MCNP based multiple-scattering corrections. This new evaluation provide a better performance on aforementioned criticality benchmarks. Overall, the discrepancies in the chromium cross sections in the evaluations have a large impact of about 1000~pcm (or 1\%) in the mentioned criticality calculations.

In this context, and in response to the NEA HPRL request, two measurements have been designed and performed: a time-of-flight measurement at n\_TOF-EAR1 carried out during summer 2022, and a $^{50}$Cr neutron activation measurement at the HiSPANoS facility of CNA~\cite{gomez2021research,macias2020first} performed in winter 2023, in which the Maxwellian Averaged Cross Section at $kT=30$~keV ($\MACS$) of $^{50}$Cr has been determined experimentally for the first time~\cite{perez2025neutron}. In this work we describe and discuss the n\_TOF measurement, of which preliminary results were presented in the WONDER 2023 Workshop (see Ref.~\cite{perez2024description}). The experimental set-up is described in Sec. \ref{sect_meas}, the extraction of the capture yield is detailed in Sec. \ref{sect_anal}, and the R-matrix resonance analysis and the results obtained are described in Sec. \ref{sect_reson}. Last, the conclusions of this work are summarized in Sec. \ref{sect_conc}.

\section{Measurement at n\_TOF}
\label{sect_meas}

\subsection{The n\_TOF facility at CERN}
\label{subsect_ntof}
The neutrons at n\_TOF are produced via spallation when 20~GeV/c$^2$ proton pulses from the Proton-Synchrotron (PS) accelerator of CERN impinge on a cylindrical $\sim$1 ton lead target~\cite{esposito2021design}, producing around 300~neutrons per proton. These pulses have a nominal intensity of $8\cdot10^{12}$ (\textit{Dedicated}) and $3\cdot10^{12}$ (\textit{Parasitic}) protons per pulse, with an average repetition rate of 0.8~Hz and a temporal width (RMS) of 7~ns. The produced neutrons are then partially moderated, allowing for an energy distribution ranging from meV to GeV. Afterwards, they travel towards three experimental areas: EAR1~\cite{guerrero2013performance} with a horizontal 185~m flight path, EAR2~\cite{weiss2015new} with a 19~m vertical flight path and NEAR~\cite{ferrari2022design}, located at only 3~m from the spallation target. 

The goal of the experiment was to measure the chromium neutron capture cross sections with high precision in the resonance region, between 1 and 100~keV, which requires a very good neutron energy resolution that is provided only at the EAR1 measuring station due to its long flight path. Overall, the experiment presented herein received $\sim7.5\cdot10^5~$ pulses during 6 weeks, accumulating a total of $\sim4\cdot10^{18}$ protons on target.

\subsection{Chromium and ancillary samples}
\label{subsect_samples}
More than 80\% of natural chromium is composed by $^{52}$Cr, hence this measurement required enriched samples to minimize backgrounds from all reactions occurring on isotopes others than those of interest:  $^{50}$Cr and $^{53}$Cr. Highly enriched chromium oxide Cr$_2$O$_3$ was purchased from Trace Science International, with the isotopic compositions given in Table \ref{table-cr2o3}. A small amount of Cu and Mo was detected in the $^{50}$Cr and $^{53}$Cr samples respectively, but these elements did not affect the measurement because only their strongest resonances were visible on the time-of-flight spectra and they are below the energy range of interest. Their contribution in the energy range of interest was negligible.

As neutron multiple-scattering effects are believed to be the main issues in previous experiments, the samples for this measurement were designed to be much thinner that those in all previous cases, thus minimizing these effects. Two samples were made for each isotope: a very thin one to measure between 1 and 10~keV, and a thicker one, but still thinner than those used in previous experiments, to cover the range from 10 to 100~keV with enough statistics. The physical characteristics of each sample are summarized in Table \ref{table-cr_thickness}, where the mass reported corresponds to the total amount of chromium oxide, from which the atomic thickness $n_{\text{at}}$ equivalent to the chromium atoms is calculated assuming the isotopic composition and uncertainty provided by the supplier of the material.

\begin{figure}
\centering
\includegraphics[width=0.9\linewidth]{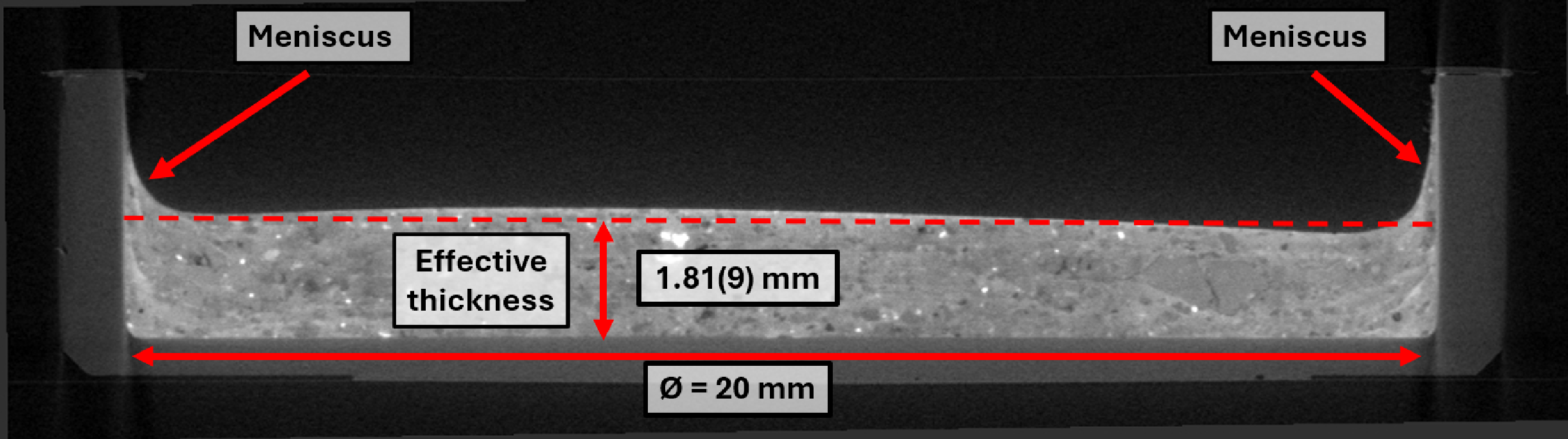}
\caption{\small X-ray Computed Tomography (CT) profile of the $^{50}$Cr-thick sample, showing a meniscus and some variation in the thickness. The dashed line indicates the average thickness.}
\label{plot-CT}
\end{figure}

The powder was pressed into a pellet inside a 20~mm diameter, 0.5~mm thick and 4.5~mm tall (or 11.5~mm for the $^{53}$Cr-thick sample) PEEK capsule with the top closed by a thin Mylar layer. This design minimizes neutron capture in the capsule, allows for a regular visual inspection of the pressed pellet to verify its integrity (as it needed to be moved in and out of the beam several times during the irradiation), and avoids losing material if the pellet would break. A series of X-ray Computed Tomographies (CT) were performed by the MME Group at CERN to each chromium sample to measure their thickness (expressed in mm in Table \ref{table-cr_thickness}) and identify any significant inhomogeneity. The CT of the $^{50}$Cr-thick sample is shown in Fig. \ref{plot-CT} as an example, where local variations of the thickness and a slightly concave meniscus at its edges can be observed. Monte Carlo simulations have shown that these aspects did not affect significantly the extraction of the capture cross section~\cite{perez2024description}. The samples were placed in the centre of a 50~mm diameter aluminium ring, glued on a Mylar foil, which is the conventional sample holder for neutron capture experiments at n\_TOF EAR1. The $^{53}$Cr-thick sample mounted in the measuring position is shown in Fig. \ref{plot-setup}.

\begin{table}
\centering
\caption{\small Isotopic composition of the enriched chromium oxide used for the measurement.}
\begin{tabular}{ c  c c c c }
\toprule
\multicolumn{1}{c}{\multirow{2}{*}{Sample}} & \multicolumn{4}{c}{Enrichment (\%)} \\ \cmidrule{2-5} 
\multicolumn{1}{c}{}  & $^{50}$Cr     & $^{52}$Cr    & $^{53}$Cr  & $^{54}$Cr \\ \midrule
                         $^{50}$Cr$_2$O$_3$  & 94.6(4) & 4.92   & 0.4  & 0.08   \\
                         $^{53}$Cr$_2$O$_3$  & 0.03     & 2.19   & 97.7(2) & 0.08 \\ \botrule
\end{tabular}
\label{table-cr2o3}
\end{table}

\begin{table}
\centering
\caption{\small Physical properties of the four chromium samples (see text for more details).}
\begin{tabular}{ l c c c }
\toprule
	   Sample & Mass (mg) & Thickness (mm) & $n_{\text{at}}$($10^{-3}$at/barn) \\ \midrule
      $^{50}$Cr-thin & 247(1) & 0.62(3) & 0.640(3) \\
      $^{50}$Cr-thick & 723(3) & 1.81(9) & 1.873(7) \\
      $^{53}$Cr-thin & 479(1) & 1.61(5) & 1.194(2) \\
      $^{53}$Cr-thick & 2362(5) & 6.84(24) & 5.885(12) \\ \botrule
\end{tabular}
\label{table-cr_thickness}
\end{table}

In addition to the chromium samples, a series of ancillary measurements were performed with the following samples: 
\begin{itemize}
    \item the two versions (4.5 and 11.5~mm tall) of the PEEK capsule without powder inside to measure the background from neutron  capture and scattering in the capsule.
    \item a natural carbon sample as a proxy for the study of the background from neutron scattering in chromium.
    \item two gold samples (20 and 80~mm diameter with 100~\textmu m thickness) for neutron energy calibration and normalization through the Saturated Resonance Method~\cite{macklin1979absolute}. 
    \item an empty aluminium ring with Mylar to determine the overall background common to all measurements, independently of the sample employed.
\end{itemize}

\subsection{Experimental set-up}
\label{subsect_setup}
\begin{figure}
\centering
\includegraphics[width=0.75\linewidth]{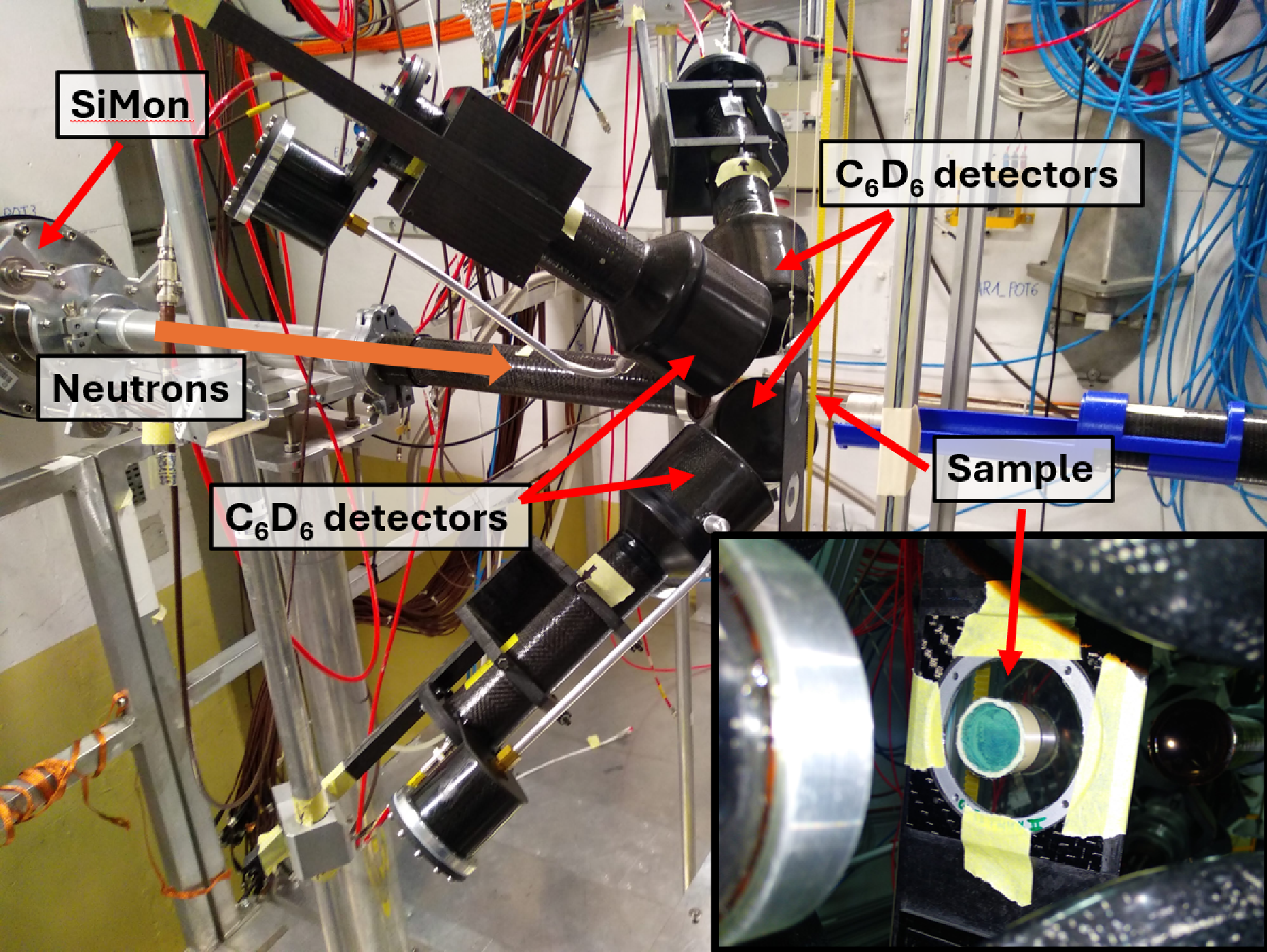}
\caption{\small Capture set-up for the chromium experimental campaign at n\_TOF EAR1, with the C$_6$D$_6$ detectors placed at 8~cm distance at a 125$^{\circ}$ backward angle with respect to the beam. Inset: the samples, like the $^{53}$Cr-thick one of the picture, are glued into a Mylar foil held by an aluminium ring.}
\label{plot-setup}
\end{figure}

The neutron capture detection set-up consisted of an array of four C$_6$D$_6$ detectors~\cite{mastinu2013new} with a cylindrical active volume of 1~L, encapsulated in a carbon fibre housing. This design by Plag et al.~\cite{plag2003optimized} reduces the neutron sensitivity of the detectors, which is needed to reduce as much as possible the background related to the large neutron scattering cross section in chromium. They also feature a very small gamma-ray detection efficiency, an attribute needed to apply the Pulse Height Weighting Technique (PHWT)~\cite{macklin1967capture} to asses the efficiency for detecting capture cascades (see Sec. \ref{subsect_TED}). The detectors were placed at 8~cm from the sample position, at a 125~degrees backward angle with respect to the neutron beam, as shown in Fig. \ref{plot-setup}, to reduce the background produced by in-beam photons undergoing Compton scattering in the sample, and to minimize possible anisotropies in the gamma-ray emission for resonances with orbital angular momentum $\ell>0$.

As it is customary at n\_TOF, the intensity of the proton pulses was monitored with a Beam Current Transformer (BCT), and the number of neutrons using the Silicon Monitor (SiMon)~\cite{marrone2004low}, which is based on the standard $^6$Li(n,\textalpha)$^3$H reaction. The ratio between the SiMon counting rate and the BCT value remained constant during the whole campaign within 2.5\%, which is then considered as the systematic uncertainty in the normalization between different runs and samples. This allows to normalize the measured counting rates as counts per nominal pulse, i.e., $8\cdot10^{12}~$protons, as given by the BCT.

The signals from all detectors and monitors were registered with the n\_TOF DAQ~\cite{abbondanno2005data,masi2018cern}, based on SPDevices ADQ14DC flash ADC digitizers, each unit featuring 4 channels with 14-bit resolution and 1~GHz sample rate. The digitized signals were stored temporally in local computers, and then sent to the CERN Tape Archive~\cite{mascetti2015disk}. They were processed by a Pulse Shape Analysis (PSA) routine specifically developed for n\_TOF data, with enough versatility to adapt to the particularities of each type of detector~\cite{vzugec2016pulse}. As a result, all the information about each signal like amplitude or time is stored in ROOT~\cite{antcheva2009root} files for further processing.

\section{Determination of the capture yield}
\label{sect_anal}
The quantity aimed to be determined from the time-of-flight measurement is the capture yield, defined as the number of captures per incident neutron as a function of the neutron energy. Experimentally, the capture yield is extracted from the following expression:
\begin{equation}
\label{eq_yield}
Y(E_n) = \frac{ C_w(E_n)-B_w(E_n) }{ E_c\cdot\Phi(E_n)\cdot F_{BIF} }F_{PHWT},
\end{equation}
where $C_w$ and $B_w$ are the total and background weighted counting rates (neglecting dead-time corrections because of the low counting rates), $E_c$ the capture cascade energy (see Sec. \ref{subsect_TED}), $\Phi$ the neutron flux at EAR1, $F_{BIF}$ the fraction of the neutron flux seen by the sample, and $F_{PHWT}$ a correction factor associated with the PHWT. The analysis to obtain all these quantities is detailed in the following sections.

\subsection{Detector and time-of-flight calibration}
\label{subsect_data-red}
The detectors were calibrated once per week both in $\gamma$-ray deposited energy and resolution using radioactive sources of $^{137}$Cs, $^{88}$Y and $^{241}$Am-$^{9}$Be which emit $\gamma$-rays of 0.662, 0.898 and 1.836, and 4.438~MeV, respectively. The end point of $^{197}$Au capture cascade at 6.512~MeV was also included in order to extend the deposited energy range. These measurements were complemented with GEANT4 simulations~\cite{allison2006geant4,allison2016recent} of the C$_6$D$_6$ detectors response, in which a realistic model of the experimental area and the whole capture set-up was implemented. A small gain shift ($\sim2\%~$ over the full campaign) was observed, especially on one of the detectors, and thus the experiment was divided into six periods, each with its own energy calibration.

A detection threshold of 150~keV in deposited energy was applied to reject the low amplitude signals with a strong contribution from noise, low energy background and the afterpulses from the PMT~\cite{akchurin2007study} that were not identified as such by the PSA. The relation between the rise time and the FHWM of the signals, which is different for real signals and afterpulses, was also used as a discrimination tool.

The kinetic energy of the neutrons is determined from the time it takes them to travel from the spallation target to the sample by means of the following non-relativistic relation:
\begin{equation}
E_n = \frac{1}{2}\frac{m_n\cdot L^2}{\left(t - t_{\gamma} + L/c + t_{off}\right)^2},
\label{eq-tof2}
\end{equation}
where $m_n$ is the mass of the neutron, $L$ is the flight path length, $t$ the time stamp of the signal, $t_{\gamma}$ the time at which the relativistic particles from the beam are observed (known as $\gamma$-flash), $c$ is the speed of light and $t_{off}$ is an offset parameter related to the Resolution Function (RF) of the facility~\cite{lorusso2004time,lo2015geant4}. Because of the random moderation length of the neutrons inside the spallation target and the borated water moderator, the relation between the time-of-flight (or flight path) and the neutron energy is not univocal. This complex relationship is described by the RF, and one of its effects is to displace the energy position of the resonances. One way of taking this into account is to include the term $t_{off}$ into Eq. (\ref{eq-tof2}). 

\begin{figure}
      \includegraphics[width=0.5\textwidth]{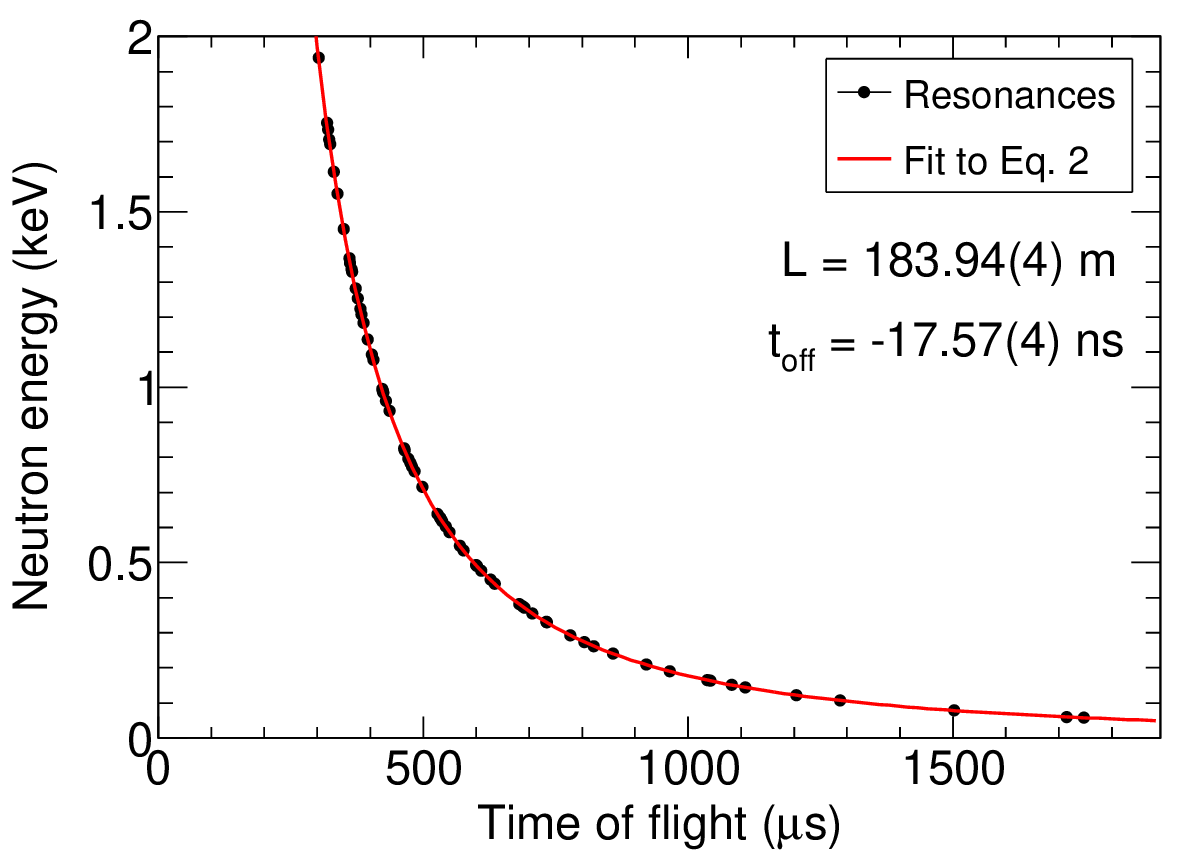}
      \includegraphics[width=0.5\textwidth]{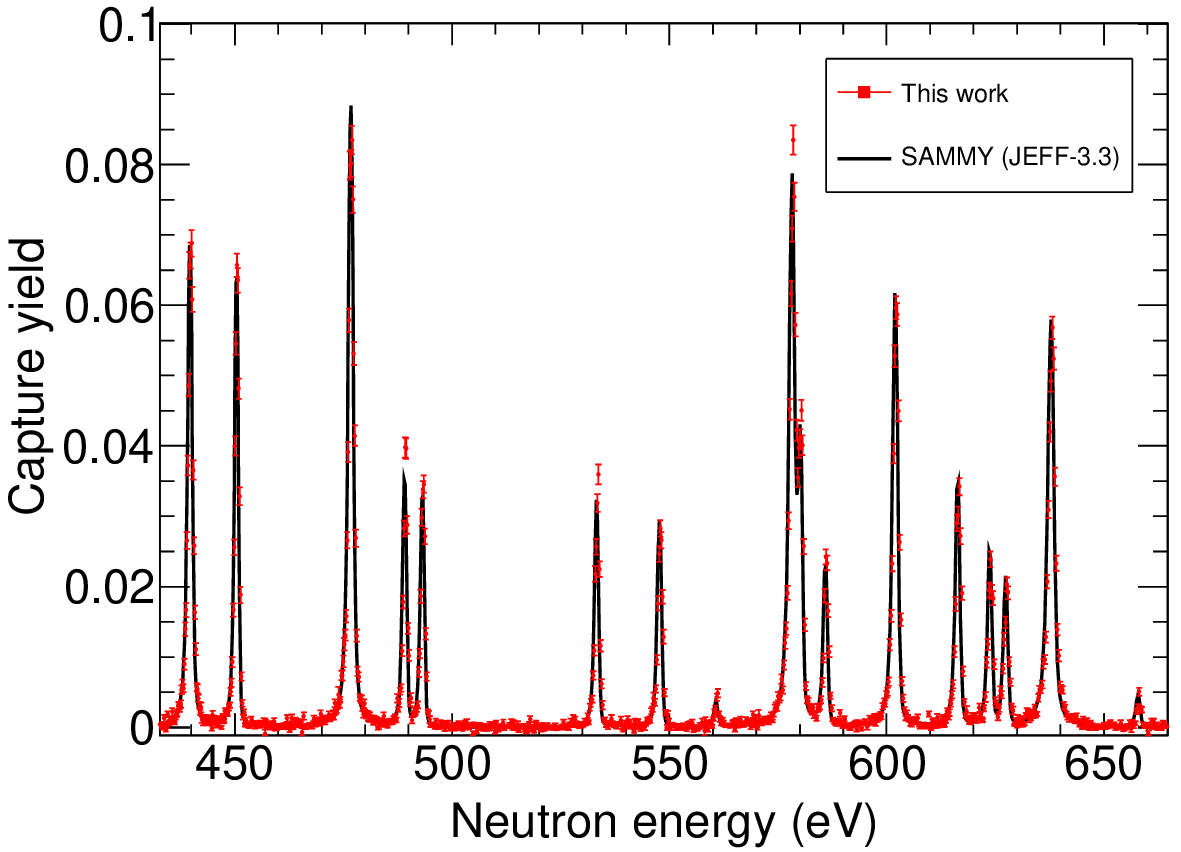}
      \caption{\small Top: Fit of the $L$ and $t_{off}$ parameters of the time-of-flight to neutron energy calibration. The points correspond to resonances of $^{197}$Au. Bottom: Illustration of the good agreement (within 0.1\%) of the $^{197}$Au resonance energies between this work and the reference JEFF-3.3 cross section.} \label{plot-au-L}
\end{figure}

The neutron energy was calibrated using the capture yield of $^{197}$Au, whose resonance energies are well known below 2~keV. The yield was analysed with the R-matrix Bayesian code SAMMY~\cite{larson2008updated} (see Sec. \ref{sect_reson}), including a numerical version of the n\_TOF RF obtained from simulations. The nominal resonance energies found in JEFF-3.3 were represented against the reconstructed values from the time-of-flight, as shown in Fig. \ref{plot-au-L}. Then, a simultaneous fit of $L$ and $t_{off}$ was performed, obtaining $L=183.94(4)~$m and $t_{off}=-17.57(4)~$ns, which allowed to reproduce the energy of the $^{197}$Au resonances within 0.1\%.

\subsection{Background determination}
\label{subsect_backg}
As mentioned in Sec. \ref{subsect_samples}, a series of measurements with auxiliary samples were performed in order to estimate the background. The different sources of background can be classified depending on whether their origin is sample-related or sample-independent.

The sample-independent background can be estimated by measuring a replica of the samples with the chromium removed. For that matter, an empty PEEK capsule (4.5 and 11.5~mm tall) was measured, and also an empty aluminium ring as dummy for measurements of gold and carbon (see below).

\begin{figure}
      \includegraphics[width=0.5\linewidth]{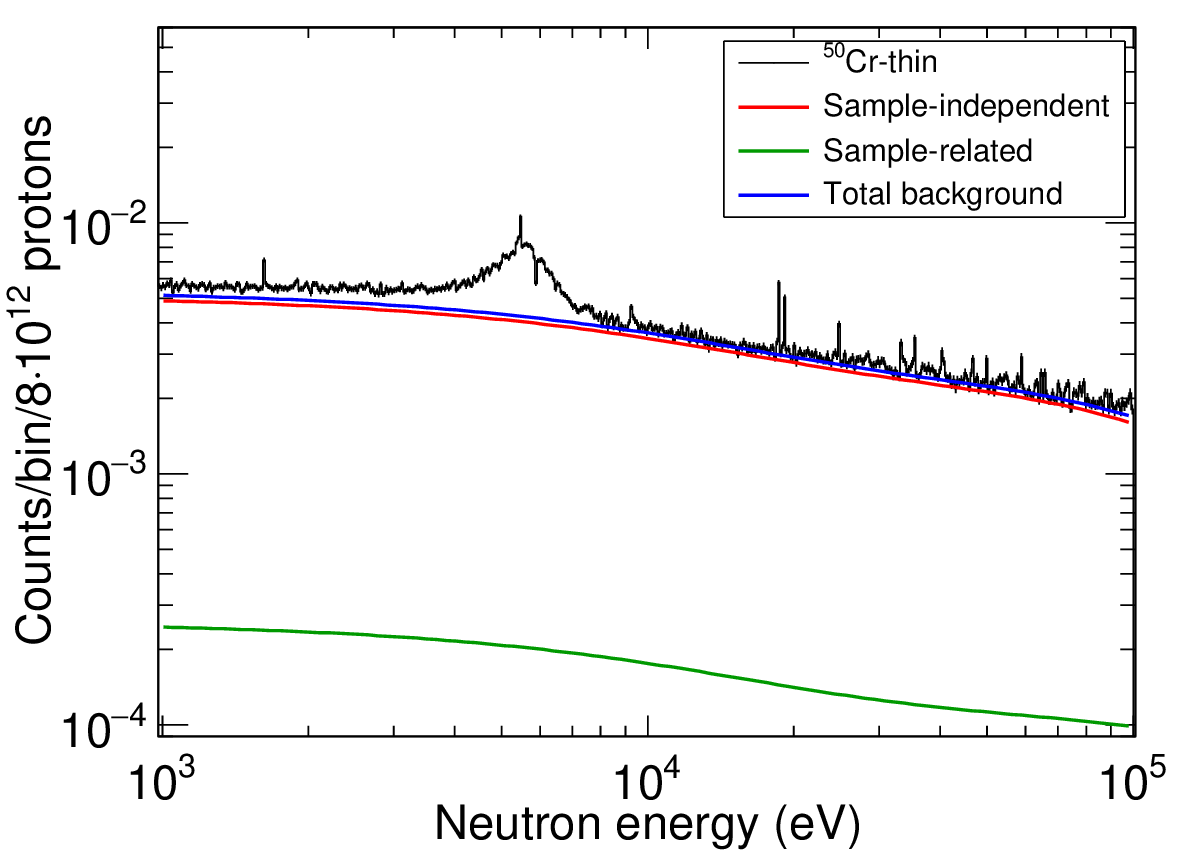}
      \includegraphics[width=0.5\linewidth]{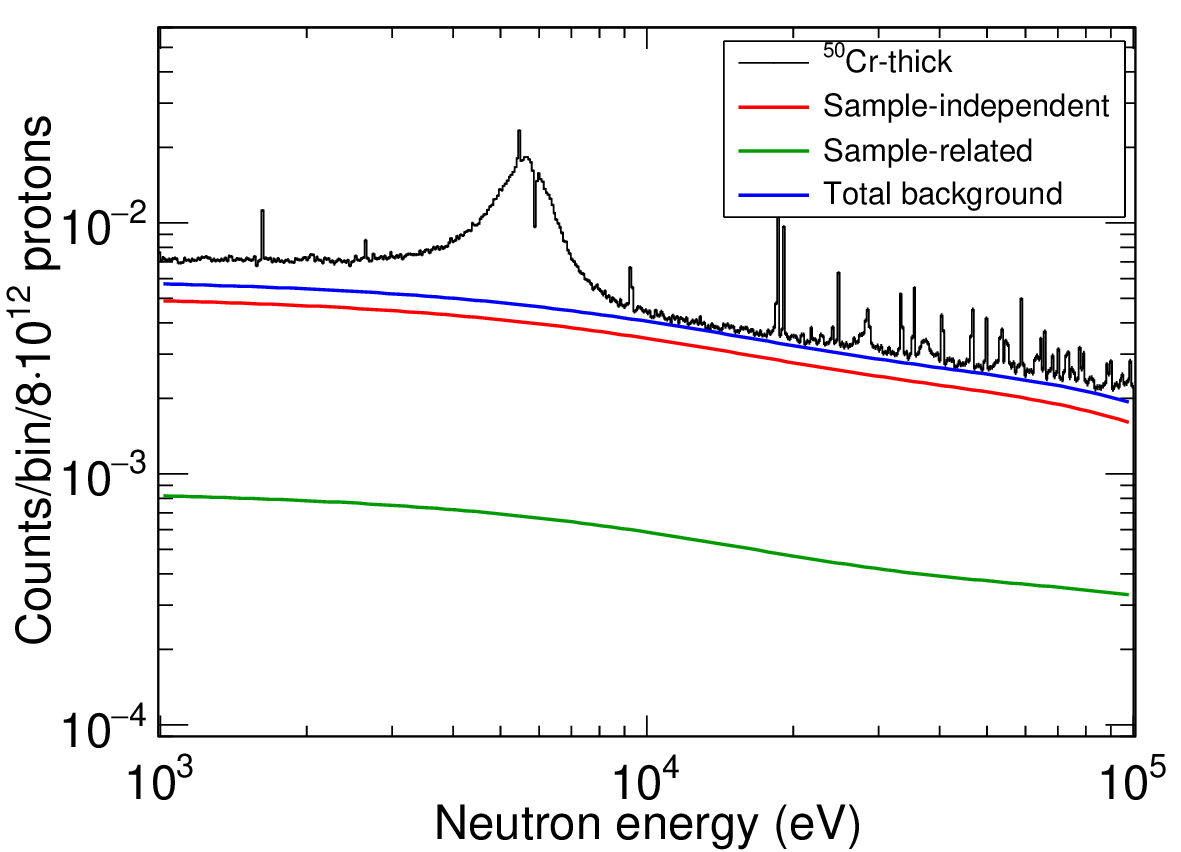}
      \includegraphics[width=0.5\linewidth]{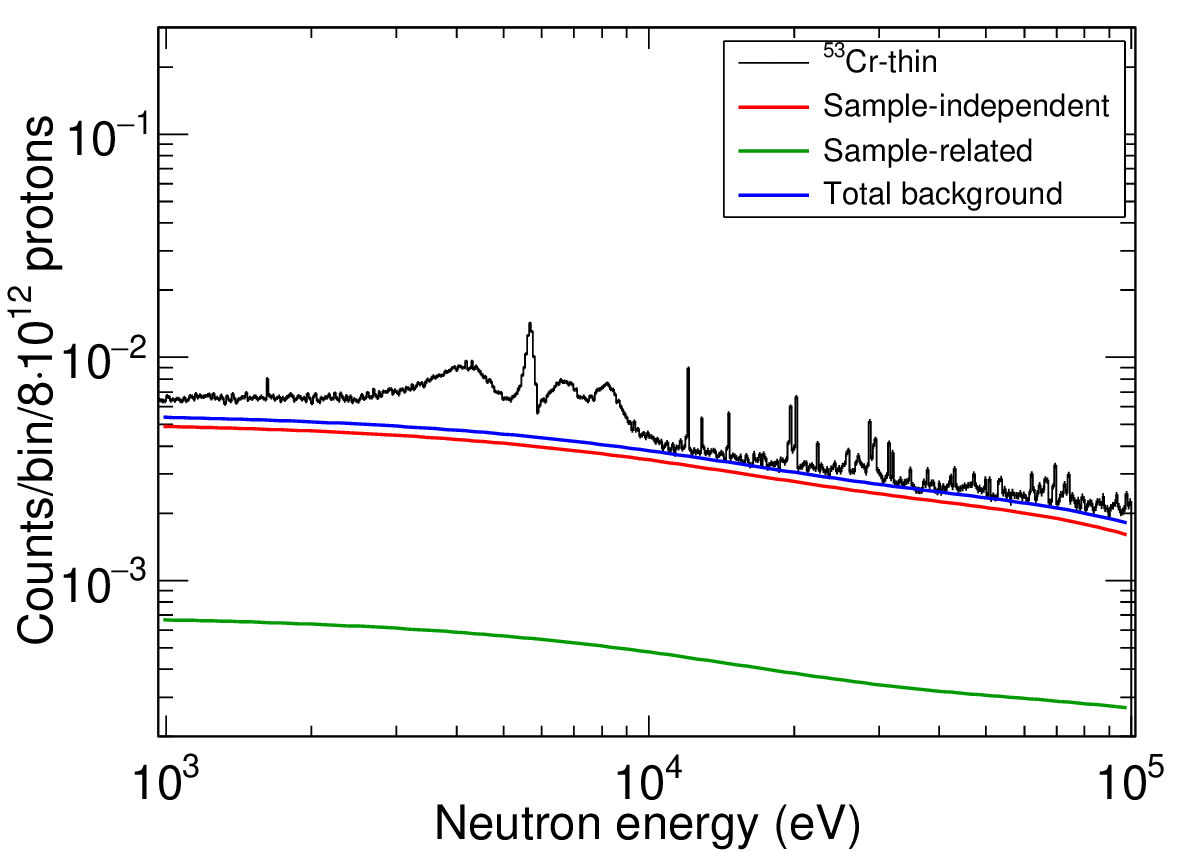}
      \includegraphics[width=0.5\linewidth]{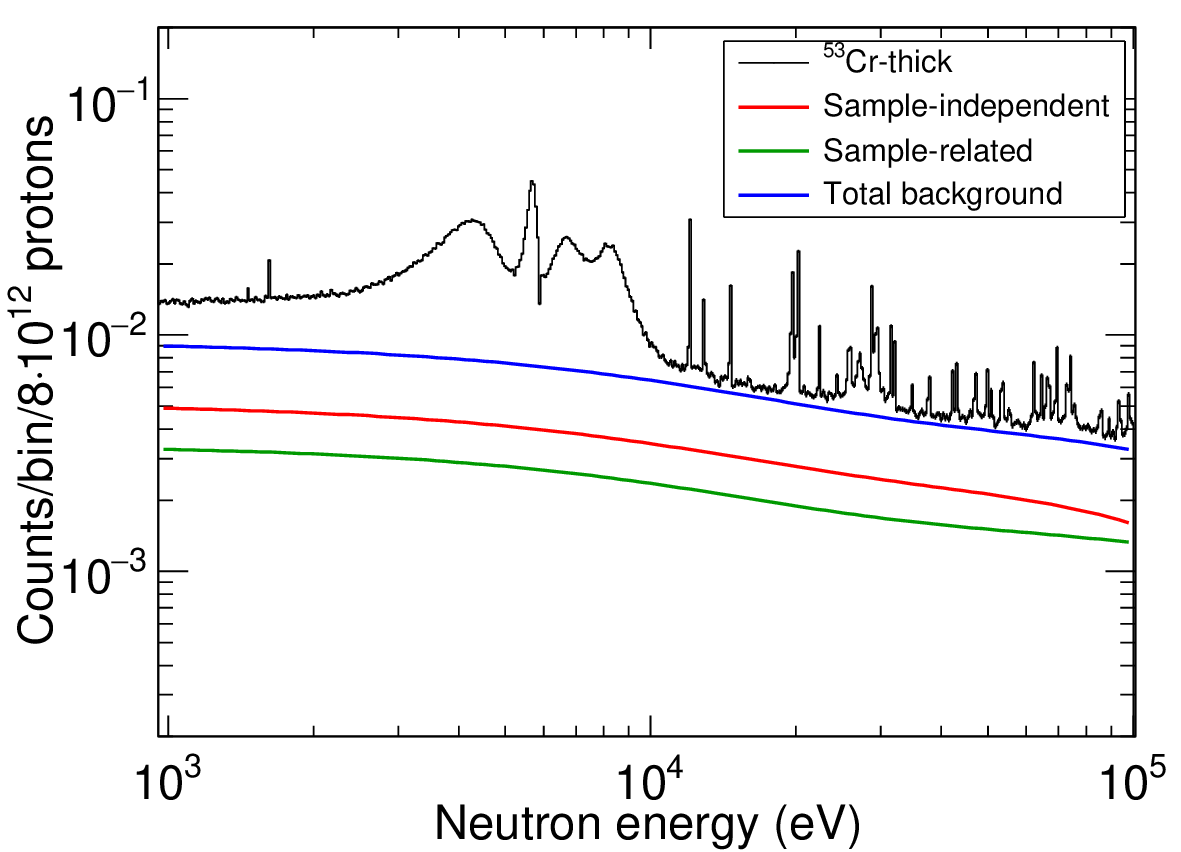}
      \caption{\small Measured spectra with the background considered for the $^{50}$Cr (top) and $^{53}$Cr (bottom). The individual background lines correspond to fits using Eq. (\ref{eq-bckg}).} \label{plot-total-bckg}
\end{figure}

The sample-related background is mainly due to neutrons scattered by the sample and then captured in its surrounding material, with a consequent $\gamma$-ray detected. Additional background could come from the scattering of in-beam $\gamma$-rays. However, it is restricted only to high Z samples and in our case it can be neglected. The sample-related background has been estimated by measuring a $^{\text{nat}}$C sample, which acts as a pure neutron scatterer because its very low capture cross section. Its low Z also allows neglecting the scattering of in-beam $\gamma$-rays. The resulting spectrum from measuring carbon has to be scaled by a factor $F_n$ to take into account the different physical characteristics of each sample and the scattering cross section that differs from chromium. This factor has been calculated as:
\begin{equation}
\label{eq-f-cnat}
F_n = \frac{n_{\text{at,Cr}}}{n_{\text{at,C}}}\Big\langle \frac{\sigma_{\text{el,Cr}}}{\sigma_{\text{el,C}}} \Big\rangle ,
\end{equation}
with $n_{\text{at}}$ the atomic thickness of each sample, and $\langle\sigma_{\text{el,Cr}}/\sigma_{\text{el,C}}\rangle$ the average ratio between Cr and C elastic cross sections in the range between 100~eV and 100~keV. In reality, $F_n$ depends on the neutron energy and its accurate calculation requires involved Monte Carlo simulations, as discussed by $\check{\text{Z}}$ugec et al.~\cite{vzugec2014geant4}. However, as shown in Fig. \ref{plot-total-bckg}, the approximation applied herein is sufficient for determining resonance parameters, since the neutron scattering background is sizeable only in the resonance valleys. The dip in the spectra near 6~keV is due to the neutron captures in the aluminium windows of the beam line. This is removed from the final capture yield so that it does not interfere with the resonance analysis.

To minimize the fluctuations due to the limited statistics on the background measurements, the energy dependence of the background spectra was parametrized using the following form:
\begin{equation}
B(E_n)=a_0+\sum_{i=1}^{n}b_i\left(1-e^{-c_iE_n}\right)e^{-d_iE_n},
\label{eq-bckg}
\end{equation}
where $a_0$, $b_i$, $c_i$ and $d_i$ are free parameters, $n=3$ for the sample-independent and $n=2$ for the sample-related background. The total counting rates of each of the four chromium samples are displayed in Fig. \ref{plot-total-bckg} together with the estimated backgrounds.
The total background is very close to the counts in the valleys between resonances. The sample-related component (green line) is then small, except for the $^{53}$Cr-thick sample. In this case, the background is still significant at the tail of the strong \textit{s}-wave resonances below 10~keV, but the cross section in this region is determined from the $^{53}$Cr-thin sample.
 
The total background depicted in Fig. \ref{plot-total-bckg} is thus considered a good estimation and hence was subtracted. The remaining smooth background component is included in the resonance analysis with SAMMY~\cite{larson2008updated} (see Sec. \ref{sect_reson}).

\subsection{Total Energy Detection technique}
\label{subsect_TED}
In order to eliminate the dependency of detection efficiency to different decay patterns from different resonances, the Total Energy Detection (TED) technique~\cite{macklin1967capture,abbondanno2004new} is used in this work. This treatment requires that the efficiency of the $\gamma$-ray detection system $\varepsilon_{\gamma}$ is \textit{i)} low enough so that as most one photon of the capture cascade is detected, and \textit{ii)} proportional to the energy of the $\gamma$-ray, $\varepsilon_{\gamma}=k\cdot E_{\gamma}$. If the proportionality factor $k=1$, the efficiency for detecting the cascade is $\varepsilon_c=E_c=S_n+\frac{A}{A+1}E_n$, with $E_c$ the energy of the cascade, $S_n$ the neutron separation energy, $A$ the atomic mass of the  target nucleus and $E_n$ the energy of the captured neutron. This means that $\varepsilon_c$ becomes numerically equal to the cascade energy, and therefore, independent of the de-excitation path of each cascade.

The C$_6$D$_6$ detectors described herein do not fulfil the second condition. However, this can be bypassed by using the Pulse Height Weighting Technique (PHWT)~\cite{macklin1967capture,tain2002accuracy}. This technique is based on the weighting of each detected signal $E_{dep}$ by an energy dependent Weighting Function (WF) in such a way that $\varepsilon_{\gamma}=E_{\gamma}$. The WF was obtained for each sample by simulating the response of the detectors to mono-energetic $\gamma$-rays, using a realistic GEANT4 model of the experimental area and the capture set-up~\cite{lerendegui2016geant4}. A total of $10^6$ events of 56 mono-energetic $\gamma$-rays between 50~keV and 10~MeV have been simulated, scoring the energy deposition of each event $E_{dep}$. 

Additionally, a correction factor $F_{PHWT}$ has been estimated to account for the overall effect due to:
\begin{itemize}
    \item the fraction of non-detected $\gamma$-rays because of the detection threshold of $E_{dep}=150~$keV;
    \item the (small) possibility to detect more than one $\gamma$-ray per cascade in a given detector (multiple-counting), even when the detection efficiency is small (e.g. 3\% for the 662~keV $\gamma$-ray emission from $^{137}$Cs);
    \item the possible emission of electrons from internal conversion instead of $\gamma$-rays.
\end{itemize}
These effects have been considered simultaneously~\cite{abbondanno2004new}, estimating $F_{PHWT}$ by simulating the response of the detectors not to individual $\gamma$-rays but to capture cascades emitted from each sample. This requires a set of realistic capture cascades, which have been obtained with the cascade generator software NuDEX~\cite{mendoza2020nudex}. NuDEX takes the level scheme, Photon Strength Functions and the branching ratios below a certain excitation energy from ENSDF~\cite{ensdfweb} and RIPL-3~\cite{capote2009ripl}. At higher energies, where the levels and branching ratios are not known, they are randomly generated according to statistical models of the nuclear level density. Each set of levels and branching ratios between all of them but the initial state is called a \textit{realization}. In practice, the de-excitation path starts at a specific resonance, and the branching of individual resonances differs. The specific choice of this branching for a fixed realization is called a \textit{subrealization}.

$F_{PHWT}$ is estimated by obtaining the deviation of the weighted response from $N_c$ simulated cascades (of the same subrealization) compared to the expected value without considering any of the effects listed above, that is:
\begin{equation}
\label{eq-F_WF}
F_{PHWT} = \frac{\sum\nolimits_i^{N_c}\sum\nolimits_jW_jR_{ij}^c}{N_cE_c},
\end{equation}
where $R_{ij}^c$ is the discretized detector response to the cascade $i$ of energy $E_c$ weighted by $W_j$, and $j$ is the number of bins of $R_{ij}^c$. The values of $F_{PHWT}$ for each sample and detector are summarized in Table \ref{table-wf-corr}. Eq. (\ref{eq-F_WF}) has also been used to estimate the systematic uncertainty associated to the WF accuracy, given by its deviation from the unity when considering the full simulated response of the cascades, without any detection threshold. By doing so for the $^{197}$Au and chromium cascades, an average value of 1.7\% has been considered as the uncertainty of the WF.

\begin{table}
\centering
\caption{\small Correction factors $F_{PHWT}$ for each individual sample and detector. The systematic uncertainty of the individual values is 1.7\% (see text for details).}
\begin{tabular}{ l cccc }
\toprule
      & \multicolumn{3}{c}{$F_{PHWT}$} \\ \midrule
	   Sample & C$_6$D$_6$\#1 & C$_6$D$_6$\#2 & C$_6$D$_6$\#3 & C$_6$D$_6$\#4 \\ \midrule
      $^{197}$Au      & 0.971 & 0.958 & 0.963 & 0.972 \\ 
      $^{50}$Cr-thin  & 0.981 & 0.978 & 0.980 & 0.979 \\ 
      $^{50}$Cr-thick & 0.971 & 0.998 & 0.972 & 0.964 \\ 
      $^{53}$Cr-thin  & 0.973 & 0.969 & 0.985 & 0.980 \\ 
      $^{53}$Cr-thick & 0.985 & 0.995 & 0.994 & 1.007 \\ \botrule
\end{tabular}
\label{table-wf-corr}
\end{table}

\begin{figure}
\includegraphics[width=0.5\textwidth]{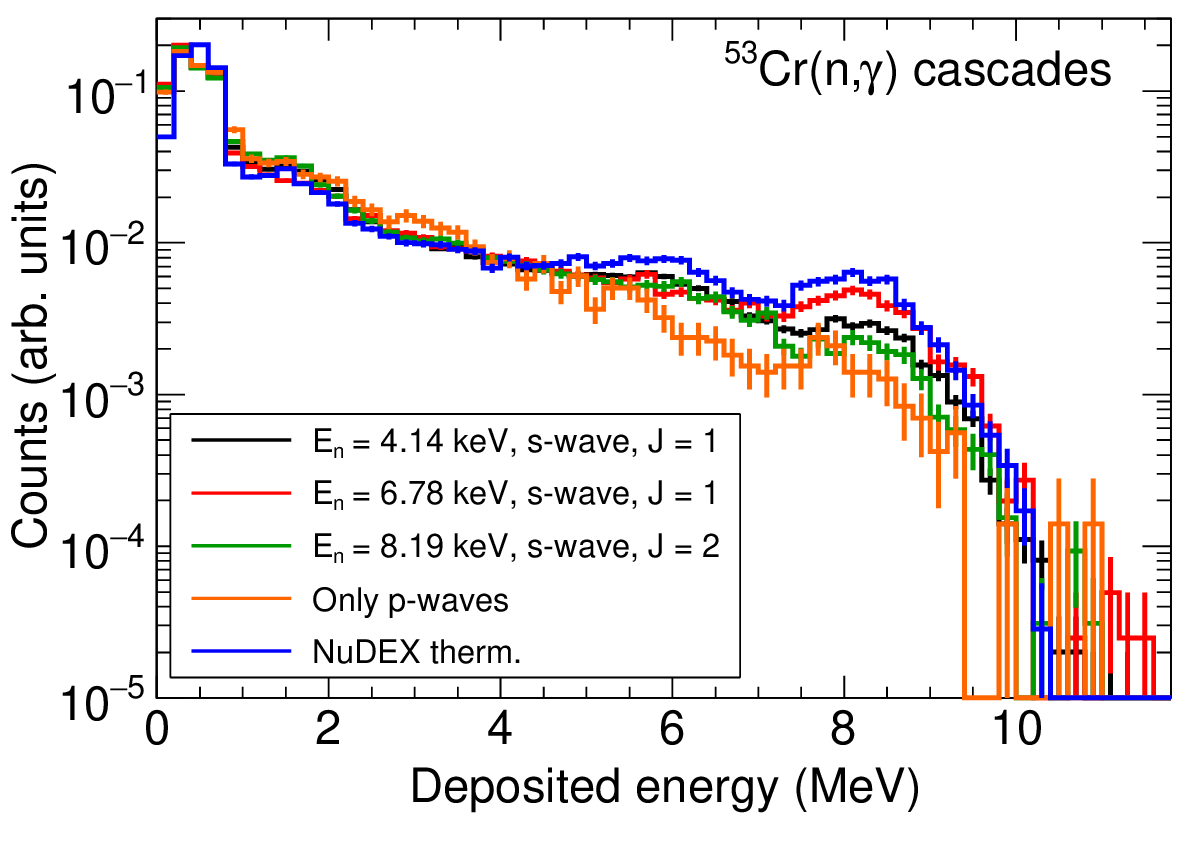}
\includegraphics[width=0.5\textwidth]{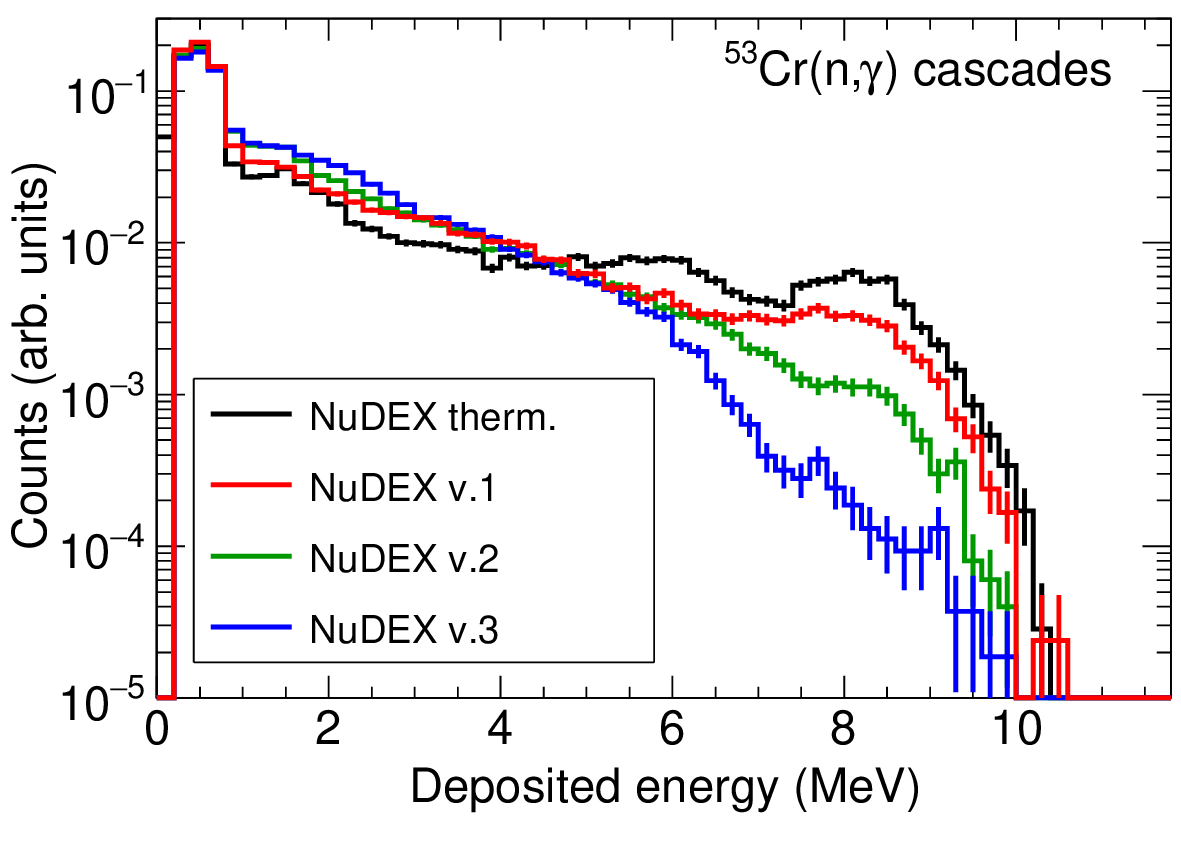}
\caption{\small Top: Deposited energy ($E_{dep}$) spectra of the three main \textit{s}-waves resonances and a sum of the cascades of multiple \textit{p}-wave resonances of $^{53}$Cr, along with a simulated thermal neutron capture cascade generated with NuDEX. Bottom: Simulated $E_{dep}$ spectra of the same thermal neutron cascade and 3 additional NuDEX subrealizations, equivalent to different resonances.}
\label{plot-casc-comp}
\end{figure}

In the particular case of chromium, the determination of $F_{PHWT}$ is not as straightforward as for $^{197}$Au, because the capture cascades change significantly between resonances with the same spin and parity. This is because the level density is relatively small, and the cascade pattern is strongly determined by the \textit{primary transitions} from each resonance. Significant fluctuations can be actually seen in the experimental deposited energy $E_{dep}$ spectra, due to primary transitions. This is shown in the top panel of Fig. \ref{plot-casc-comp} that  compares the measured spectra from three strong \textit{s}-wave resonances from the $^{53}$Cr-thick sample. Bottom panel of Fig. \ref{plot-casc-comp} then provides the simulated spectra corresponding to different NuDEX subrealizations (within one realization), showing expected differences in spectra at least comparable to those from experiment. Accordingly, the correction factor $F_{PHWT}$ can change for each resonance, as shown by Mendoza et al.~\cite{mendoza2023neutron}. The conclusion from that study is that, keeping the detection threshold as low as 150~keV, even in the case of chromium in which the capture cascades change significantly, the effect in $F_{PHWT}$ is as small as 0.7\% for $^{50}$Cr and 0.9\% for $^{53}$Cr. Consequently, we considered an additional and conservative 1\% uncertainty affecting $F_{PHWT}$.

\subsection{Normalization through the Saturated Resonance Method}
\label{subsect_SRM}
\begin{figure}
      \centering
      \includegraphics[width=0.75\linewidth]{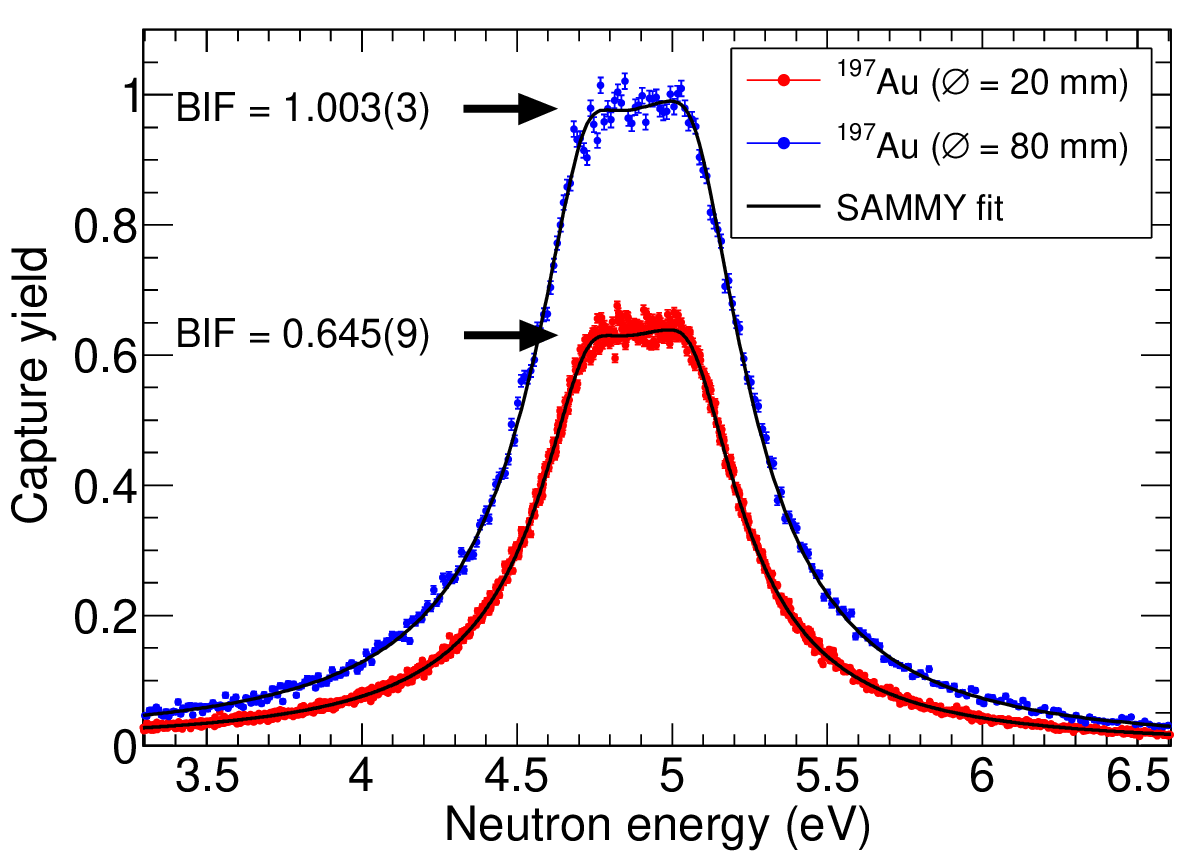}
      \caption{\small Capture yield corresponding to the 20~mm and 80~mm diameter $^{197}$Au samples fitted with SAMMY to get the Beam Interception Factor ($F_{BIF}$).} \label{plot-srm}
\end{figure}

The Beam Interception Factor (BIF) is defined as the fraction of neutrons in the beam seen by the sample, which is introduced in the calculation of the capture yield as a normalization factor $F_{BIF}$. This factor is obtained through the so-called Saturated Resonance Method (SRM)~\cite{macklin1979absolute} using the 4.9~eV resonance of $^{197}$Au measured with a thick target and thus featuring a saturation of the yield, demonstrated as a flat resonance top, as all neutrons impinging the sample are captured (see Fig. \ref{plot-srm}). The capture yield of the 20~mm $^{197}$Au sample has been obtained for each detector and analysed with SAMMY~\cite{larson2008updated} leaving the normalization factor free to vary. The resulting normalization factors provide the values of $F_{BIF}$, which agree within 1.4\% for individual detectors, yielding an average value of $F_{BIF}=0.645(9)$. The same procedure was performed to the capture yield of the 80~mm diameter gold sample, which is larger than the beam diameter, obtaining 1.003(3), in a perfect agreement with the expected value of 1. 

The $F_{BIF}$ value for the chromium samples is then considered as that of the 20~mm gold sample because they feature the same diameter. However, this requires the chromium and gold samples to be perfectly aligned with respect to each other, something that is achieved within 0.5~mm using an alignment laser system. Adopting the description of the beam profile from Guerrero et al.~\cite{guerrero2013performance}, the associated uncertainty in $F_{BIF}$ is conservatively estimated as 2\%.

\subsection{Capture yield: PWHT implementation for $^{50,53}$Cr}
\label{subsect_PWHT}
The implementation of the PHWT enhances the statistical fluctuations of the data because the weights are assigned in a \textit{signal-by-signal} basis, something that has been studied in detail recently by Mendoza et al.~\cite{mendoza2023neutron}. Having limited statistics when applying the PHWT is equivalent to sampling poorly the detector response. This becomes critical when the statistics are very limited, causing the overall weighting to vary significantly between neighbouring bins. As a consequence, fluctuations on individual points are enhanced, inducing deformations in the shape of the resonances after the weighting process. In our case, this issue is severely affecting the yield above $\sim$20~keV due to the limited statistics in combination with a significant contribution of high-energy $\gamma$-rays to chromium spectra.

\begin{figure}
      \centering
      \includegraphics[width=0.75\linewidth]{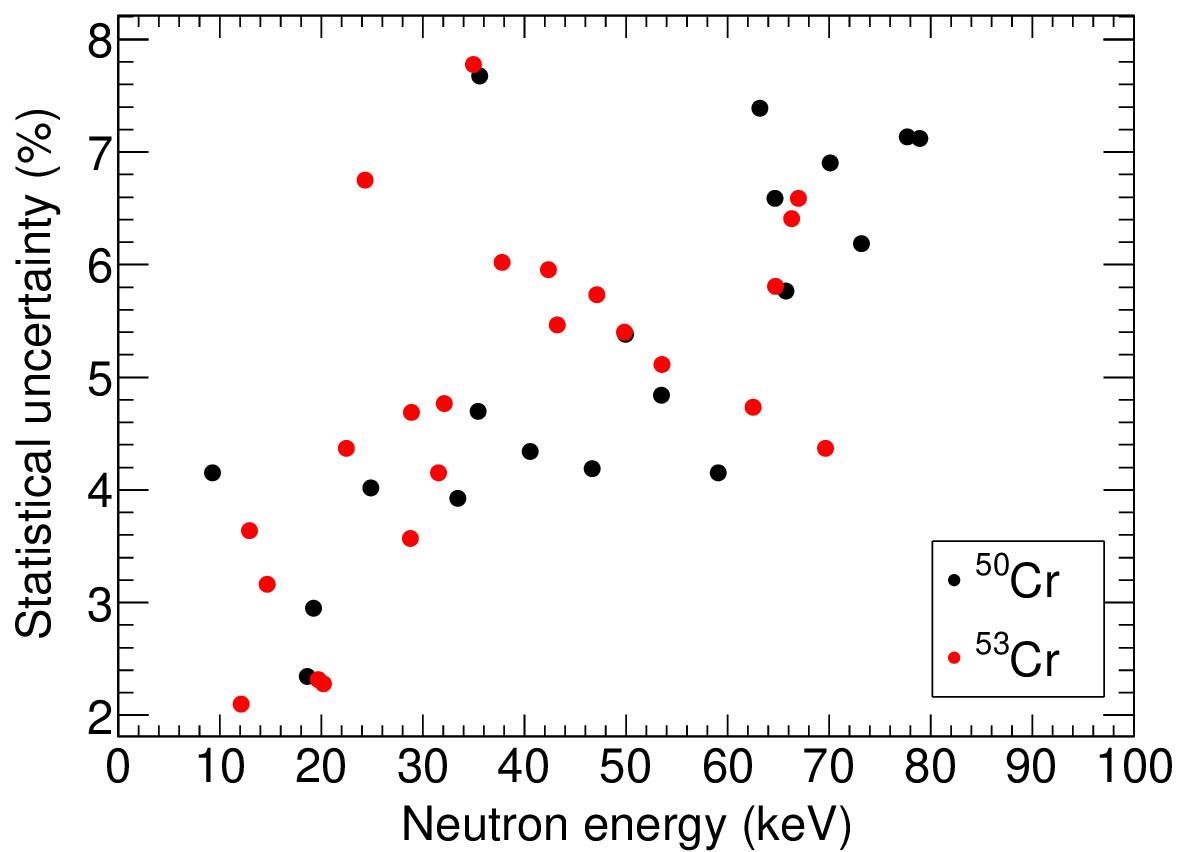}
      \caption{\small Statistical uncertainties extracted from the resonance area of $^{50}$Cr and $^{53}$Cr, corresponding to the uncertainty of each $WF_{res}$. These were considered in the overall uncertainty of the final capture yield.} \label{plot-stat-unc}
\end{figure}

As shown with detail in Ref.~\cite{PerezMaroto:2920339}, we can suppress the fluctuations in weighted counts $C_w$ by defining energy regions where the detection efficiency is constant. We have developed a new method in which, instead of applying directly the PHWT calculated in narrow neutron energy bins, we determine a weighting factor $WF_{res}$ independently for each resonance: i.e. a \textit{resonance weighting factor} (RWF) is applied. The $WF_{res}$ factors are obtained by computing the ratio between the area of each resonance in the unweighted and weighted yields. This method is a generalization of the Average Weighting Factor (AWF) technique introduced in Ref.~\cite{lerendegui2018radiative}. Our method is useful for nuclei where the detection efficiency changes significantly between individual resonances, as in the case of chromium, so it works very well for isolated resonances. Obviously, the RWF can only be calculated for resonances with enough statistics so $WF_{res}$ can be accurately extracted, and otherwise the AWF had to be used instead, as it is done in the valleys between resonances. We have included an additional systematic uncertainty in the capture yield, corresponding to the statistics of each resonance as it affects the $WF_{res}$. This is shown in Fig. \ref{plot-stat-unc}. Since the statistics of the main \textit{s}-wave resonances located between 1 and 10~keV are much higher, they are not significantly affected by the fluctuations, so the yield in that region has been obtained applying the "standard" PHWT, and thus is not affected by this additional uncertainty.

To summarize:
\begin{itemize}
    \item The capture yield between 1 and 10~keV, which is the most critical region, has been extracted applying the standard PHWT to the thin samples measurements.
    \item For the range between 10 and 100~keV, the capture yield has been extracted from the thick samples. Parameters of 42 resonances have been analysed using the new RWF technique, and the remaining ones (with poor statistics) along with the data between resonances have been analysed by the AWF technique. The capture yield in this region has an additional systematic uncertainty because of this procedure (see Table \ref{table-unc}). 
\end{itemize}

\subsection{Systematic uncertainty}
The different sources of systematic uncertainties have been described throughout the text, and are summarized in Table \ref{table-unc}. The estimation of uncertainties is conservative but still yields an overall accuracy between 5\% and 9\% in the energy range of interest (1-100~keV). This is indeed within the 8-10\% requested in the NEA High Priority Request List (HPRL) motivating this experiment.

\begin{table}
\centering
\caption{\small Contributions to the systematic uncertainty of the capture yields. As described in Sec. \ref{subsect_PWHT}, the RWF correction applies only to neutron energies above 10~keV.}
\begin{tabular}{lcc} \toprule
\multirow{2}{*}{Contribution} & \multicolumn{2}{c}{Syst. unc. (\%)} \\ \cmidrule{2-3}
                        & $^{50}$Cr & $^{53}$Cr \\ \midrule
        Sample thickness &  0.4  & 0.2  \\
        Beam monitoring & 2.5 & 2.5 \\
        Neutron flux shape &  2    & 2    \\
        Saturated Resonance Method & 1.5 & 1.5 \\
        Sample alignment & 2 & 2 \\
        \underline{PHWT}:\\
        PHWT implementation & 1.7 & 1.7 \\
        Cr cascades $\circledast$ $F_{PHWT}$  & 1 & 1 \\
        RWF & 2.3 - 8 & 2.1 - 8 \\ \midrule
        \textbf{Overall (1 - 10~keV)} & \textbf{5} & \textbf{5} \\ 
        \textbf{Overall (10 - 100~keV)} & \textbf{5 - 9} & \textbf{5 - 9} \\ \botrule
\end{tabular}
\label{table-unc}
\end{table}

\section{Resonance analysis}
\label{sect_reson}
The resonances in the capture yield were analysed with the multilevel multichannel R-matrix code SAMMY~\cite{larson2008updated}. The code allows for a Bayesian fitting of the capture yield in the Resolved Resonance Region (RRR) using the Reich-Moore R-matrix approximation~\cite{reich1958multilevel}, including several experimental effects like the Doppler and RF broadening, the multi-isotopic composition of the samples and the self-shielding and multiple-scattering effects. With SAMMY it is also possible to include the residual background present in the yield (see Sec. \ref{subsect_backg}) by fitting the valleys between resonances.

\subsection{Resonance analysis with SAMMY}
\label{subsect_ind-res}
One of the main goals of this measurement was to minimize the multiple-scattering effects in the capture yield. For this purpose, two samples with different thickness were used for each isotope. The region with the wide \textit{s}-wave resonances between 1 and 10~keV was analysed using the very thin samples, so multiple-scattering effects are much less relevant. The rest up to 100~keV was analysed with the thick samples. 

For each resonance both neutron $\Gamma_n$ and capture $\Gamma_\gamma$  widths can be fitted; but, if both are fitted simultaneously, correlations appear and their uncertainty increases. To prevent this, whenever it was possible only one of the widths was fitted, usually $\Gamma_\gamma$ since for most resonances $\Gamma_\gamma\ll\Gamma_n$ and thus the radiative kernel is dominated by it (see Eq. (\ref{eq-kernel})). The resonance energy $E_n$ was fitted in all cases. 

The resonance parameters from JEFF-3.3~\cite{plompen2020joint} and CENDL-3.2~\cite{ge2020cendl} evaluations were used as a initial guess for the Bayesian fit, choosing the ones resulting in the best result. The spin was kept as in the evaluations, unless the fit was not satisfactory or the evaluations do not agree, in which case was changed (not fitted) to provide a better result.

A total of 33 resonances have been observed and analysed for $^{50}$Cr and 51 for $^{53}$Cr. The complete list of resonance parameters (with their correlation when applicable) and radiative kernels (see Sec. \ref{subsect_comp2}) are listed in Tables \ref{table-cr50-res} and \ref{table-cr53-res2} of the Appendix.

\subsection{Results and discussion between 1 and 10~keV}
\label{subsect_comp1}
\begin{figure}
    \includegraphics[width=0.5\linewidth]{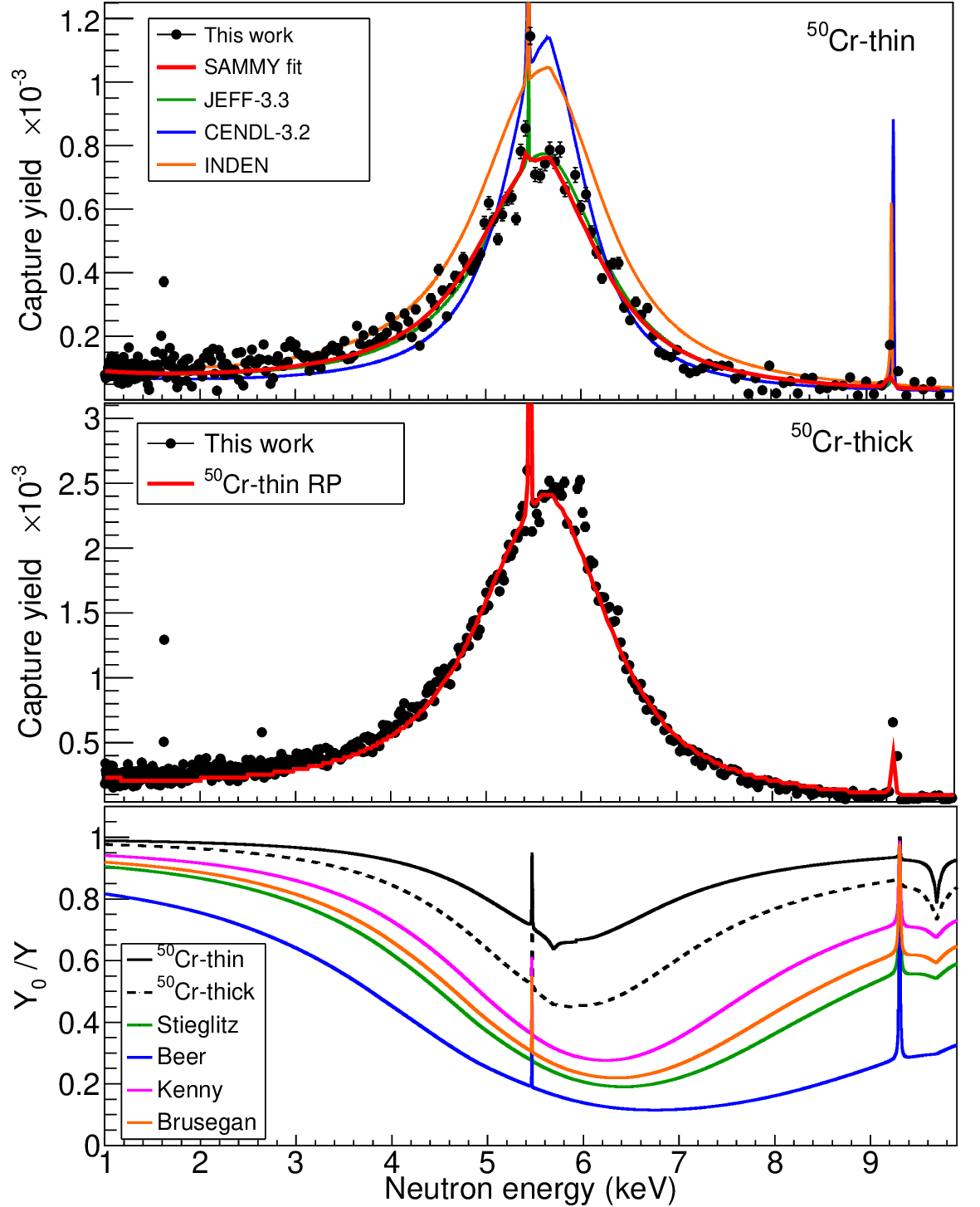}
    \includegraphics[width=0.5\linewidth]{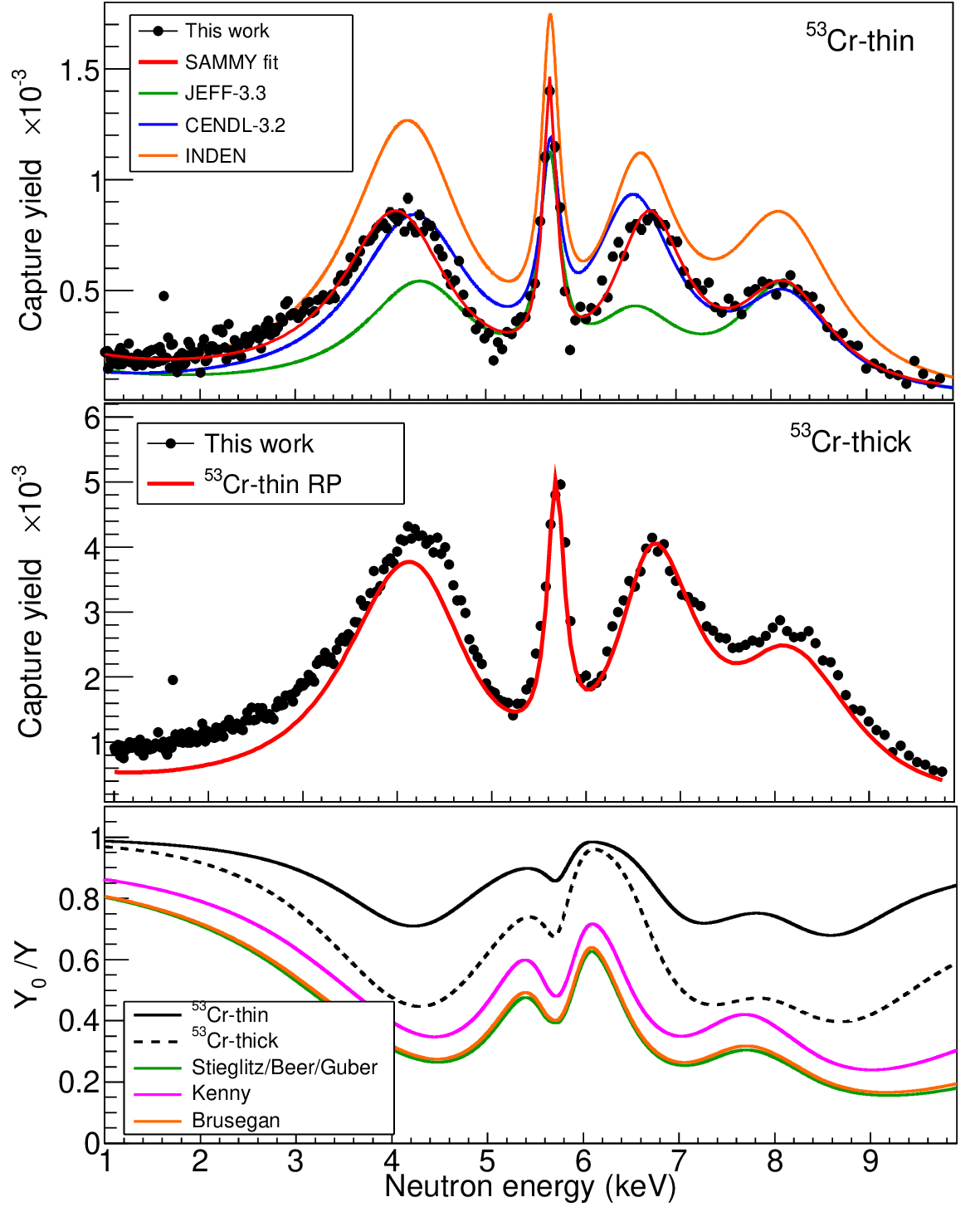}
      \caption{\small Top: Capture yield of $^{50}$Cr (left) and $^{53}$Cr (right) thin samples measured at n\_TOF and fitted with SAMMY, compared with the predictions using the parameters from evaluations. Middle: Capture yield of the thick samples compared to the prediction of SAMMY using the parameters from the thin sample fitting. Bottom: Fraction of the capture yield corresponding to the captures without scattering, for the samples used at n\_TOF samples and in the previous measurements~\cite{stieglitz1971kev,beer1975kev,kenny1977neutron,brusegan1986high,guber2011neutron}.} \label{plot-main-res}
\end{figure}

The range between 1 and 10~keV is the most important for criticality benchmarks because of a cluster of strong \textit{s}-wave resonances in both $^{50}$Cr and $^{53}$Cr. These resonances are also the main source of multiple-scattering effects, which severely affected previous measurements. The measured capture yield with the corresponding SAMMY fit are displayed and compared with the expected yields from the JEFF-3.3, CENDL-3.2 and INDEN~\cite{nobre2021newly} evaluations in the top panels of Fig. \ref{plot-main-res}. The figure illustrates very clearly the large difference between evaluations. In this particular region, our result for $^{50}$Cr is in very good agreement with JEFF-3.3 but significantly deviates from the other evaluations. Note that although the shape of the resonance is very different from CENDL-3.2, the resonance parameters are such that the resonance kernel (see Sec. \ref{subsect_kernel}) is in very good agreement with this evaluation. Regarding $^{53}$Cr, the best agreement, but surely not perfect, is reached with CENDL-3.2. Notably, for both isotopes the INDEN evaluation clearly overestimates the cross sections. The structures at 1.6~keV and 2.6~keV correspond to the first resonances of $^{52}$Cr and $^{63}$Cu impurities.

To illustrate the problems related to the use of thick samples in previous measurements (significantly thicker than all of our samples), the resonance parameters from the analysis of the thin samples has been used in SAMMY to predict the yield corresponding to our thick samples. As shown at the middle panels of Fig. \ref{plot-main-res}, the predictions do not reproduce our spectra with the parameters deduced from our thin sample, specially for the $^{53}$Cr case, evidencing the limited capability of SAMMY for estimating accurately the multiple-scattering interactions prior to the neutron capture for very thick samples. 
The total capture yield can be expressed at the sum of individual components $Y_i$ as $Y(E_n)=\sum_{i}Y_i(E_n)$, where the index $i$ indicates the number of scatterings before the capture. To better visualize the importance of the multiple-scattering effects, we display in the bottom panels of Fig. \ref{plot-main-res} the fraction of the total capture yield that corresponds to $Y_0$, i.e., the fraction of captures without any previous scattering. It is clear that the multiple-scattering components are an important contribution to the capture yield in our thick samples, contributing in some regions by more than 50\% to the yield. For illustration, the figure also shows $Y_0$ (based on our parameters from the thin samples) that would correspond to the samples used in the previous experiments~\cite{stieglitz1971kev, beer1975kev, kenny1977neutron, brusegan1986high, guber2011neutron}. These samples were 8 to 30 and 7 to 12 times thicker than our thin $^{50}$Cr and $^{53}$Cr samples, respectively.

\subsection{Results and discussion between 10 and 100~keV}
\label{subsect_comp2}
From 10~keV onwards, we used the yield of the thick samples to perform the resonance analysis, because of the much better statistics. 

\begin{figure}
      \includegraphics[width=0.5\linewidth]{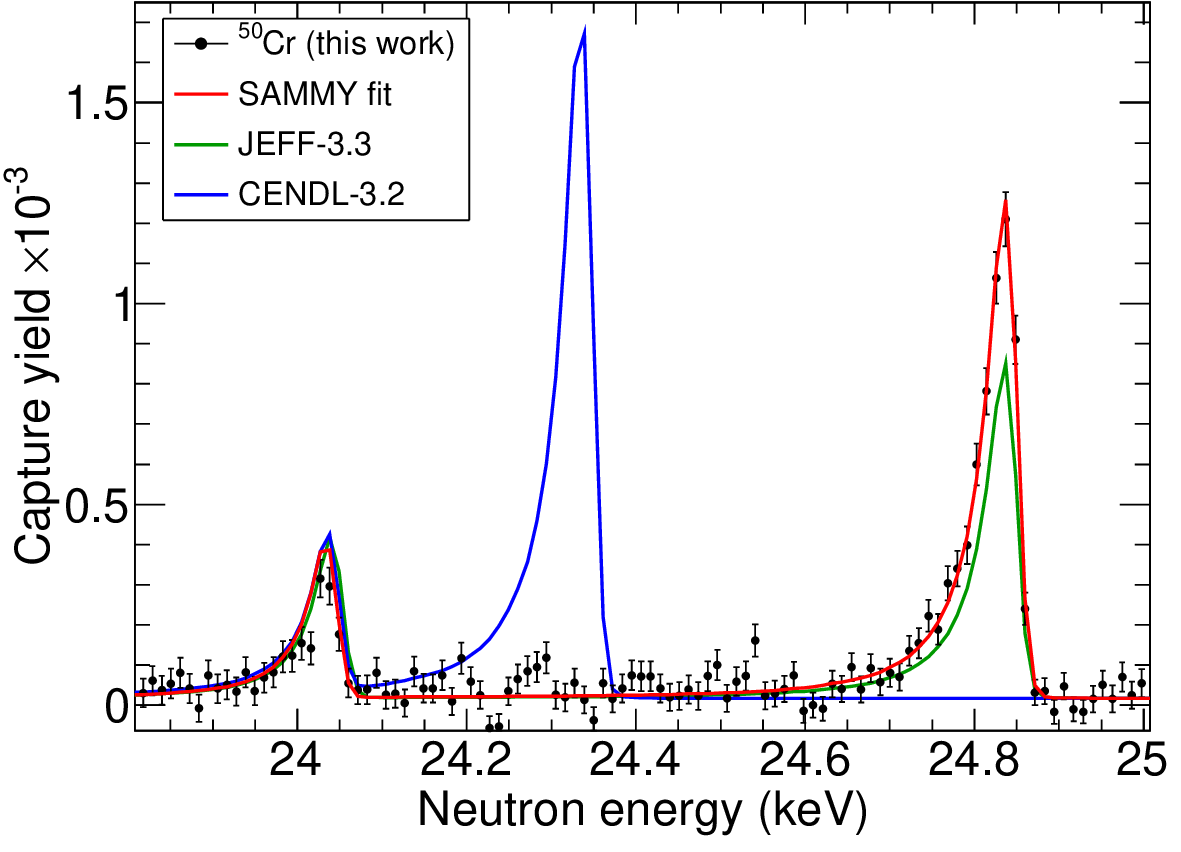}
      \includegraphics[width=0.5\linewidth]{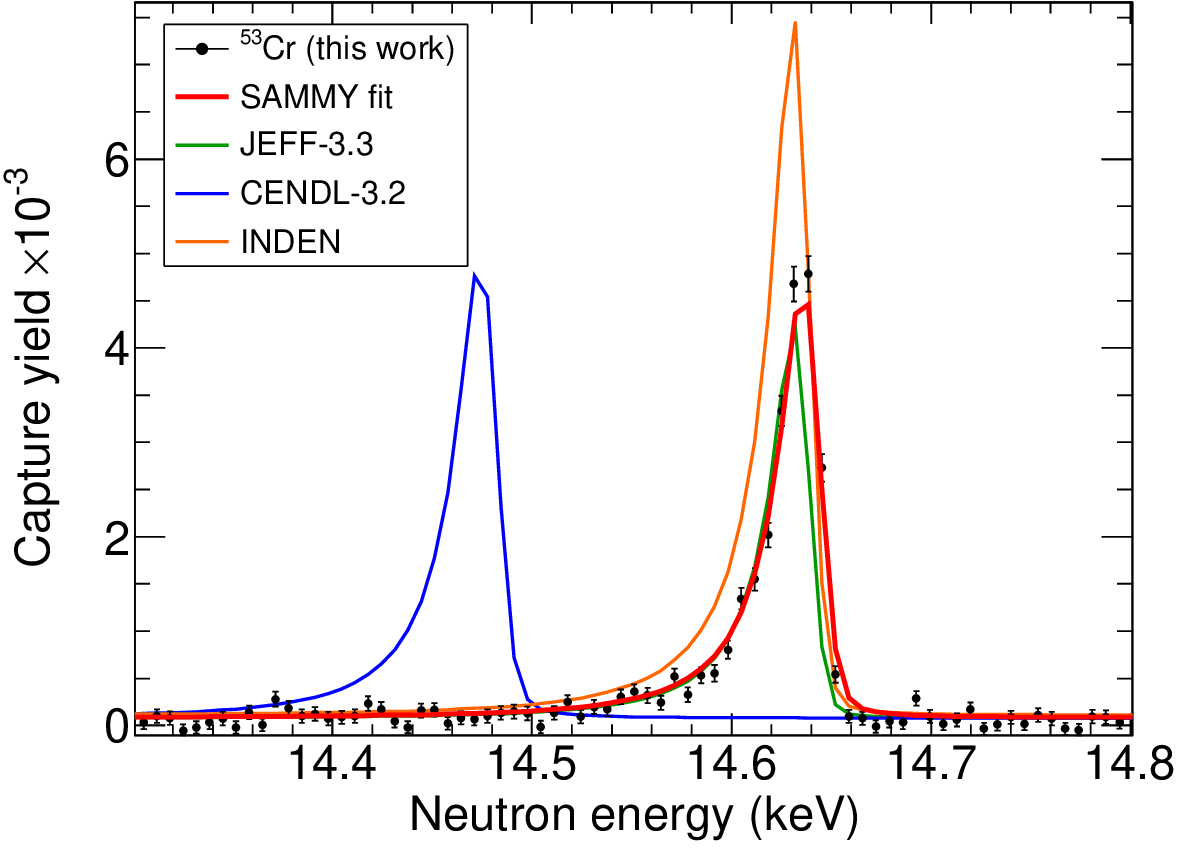}
      \includegraphics[width=0.5\linewidth]{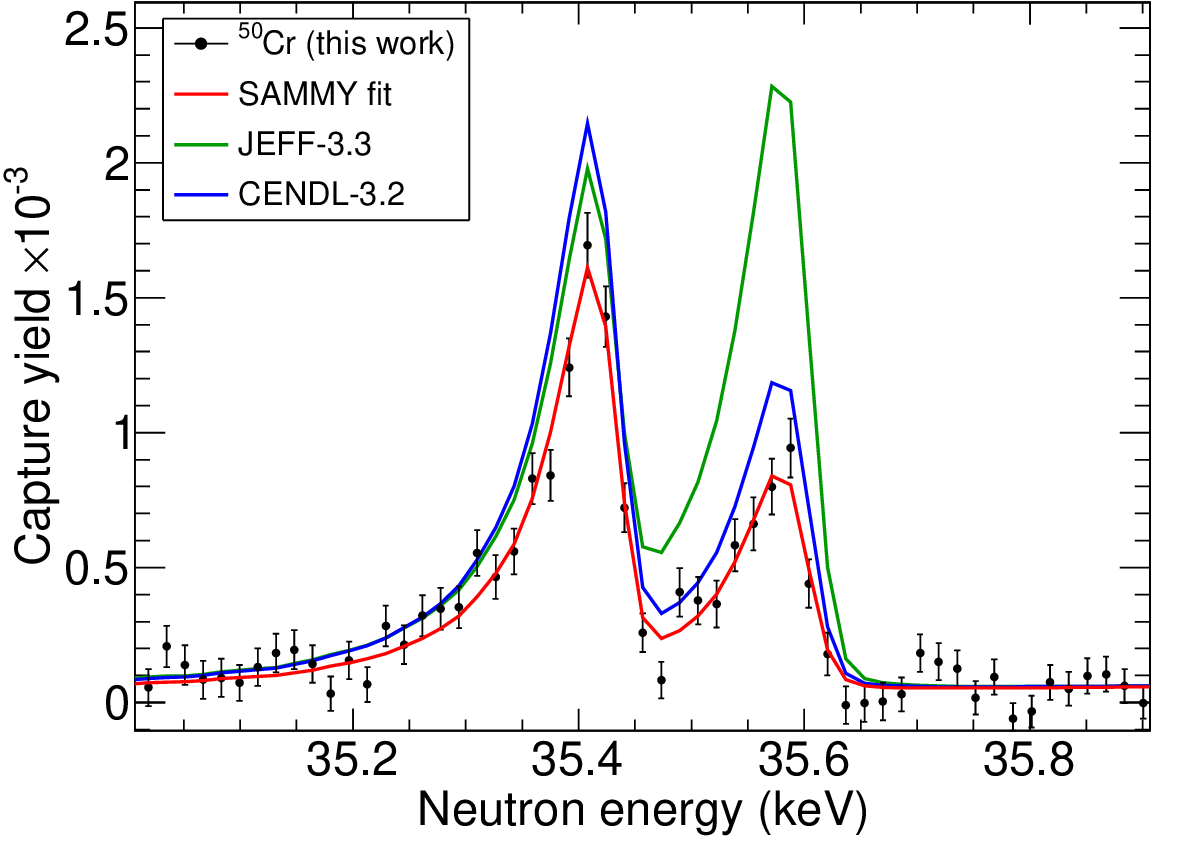}
      \includegraphics[width=0.5\linewidth]{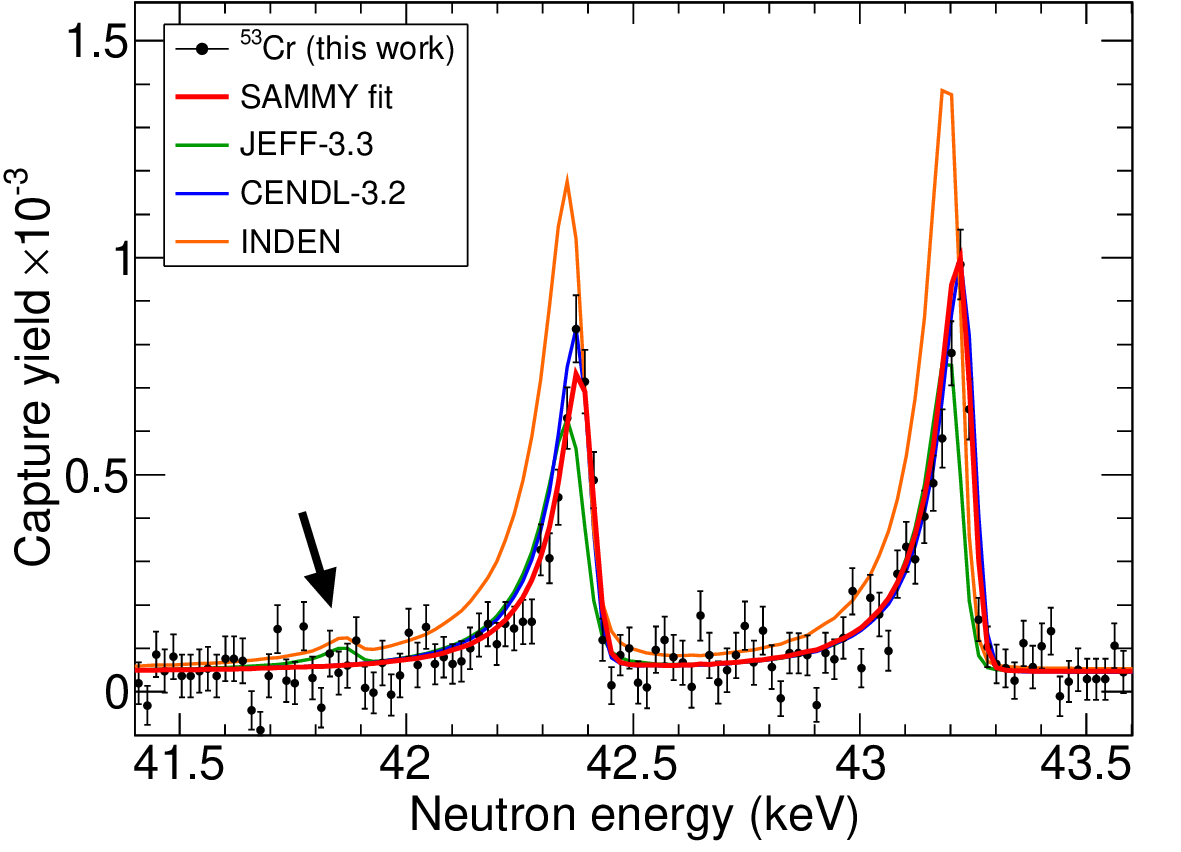}
      \includegraphics[width=0.5\linewidth]{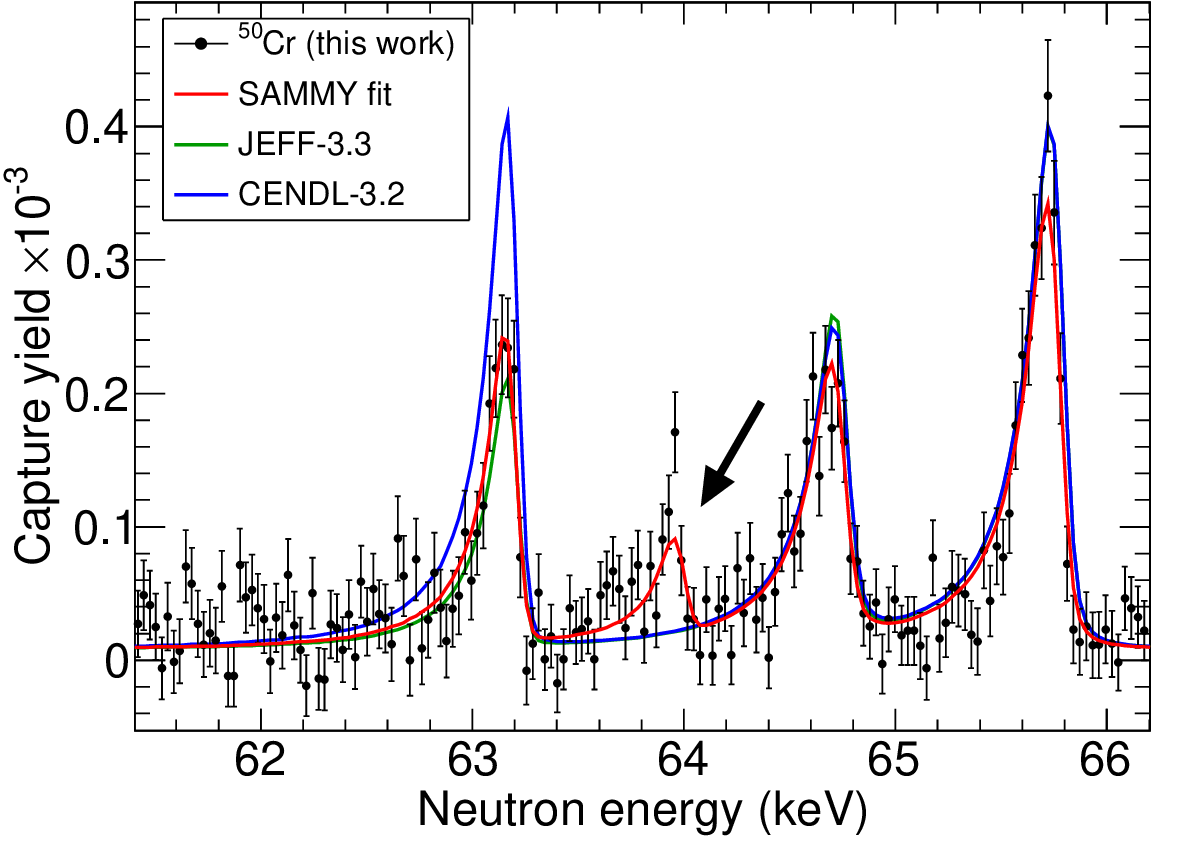}
      \includegraphics[width=0.5\linewidth]{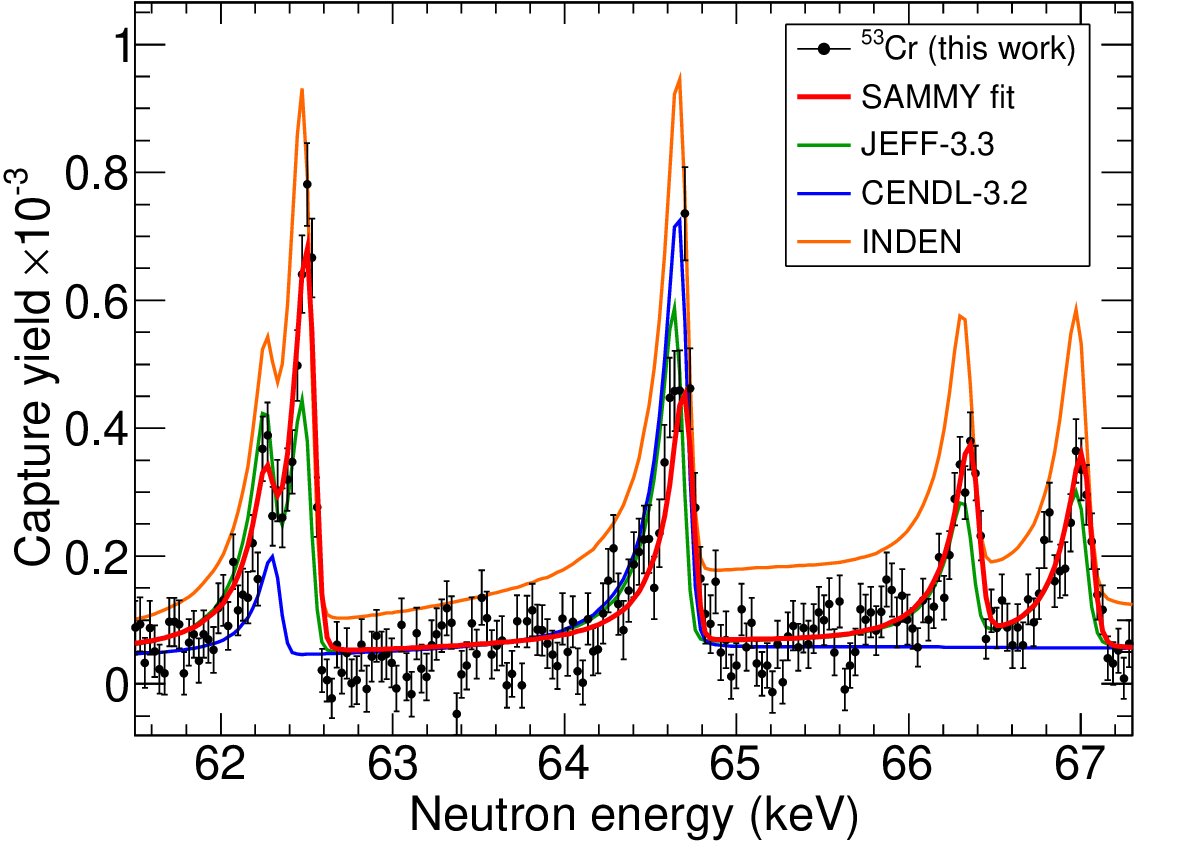}
      \includegraphics[width=0.5\linewidth]{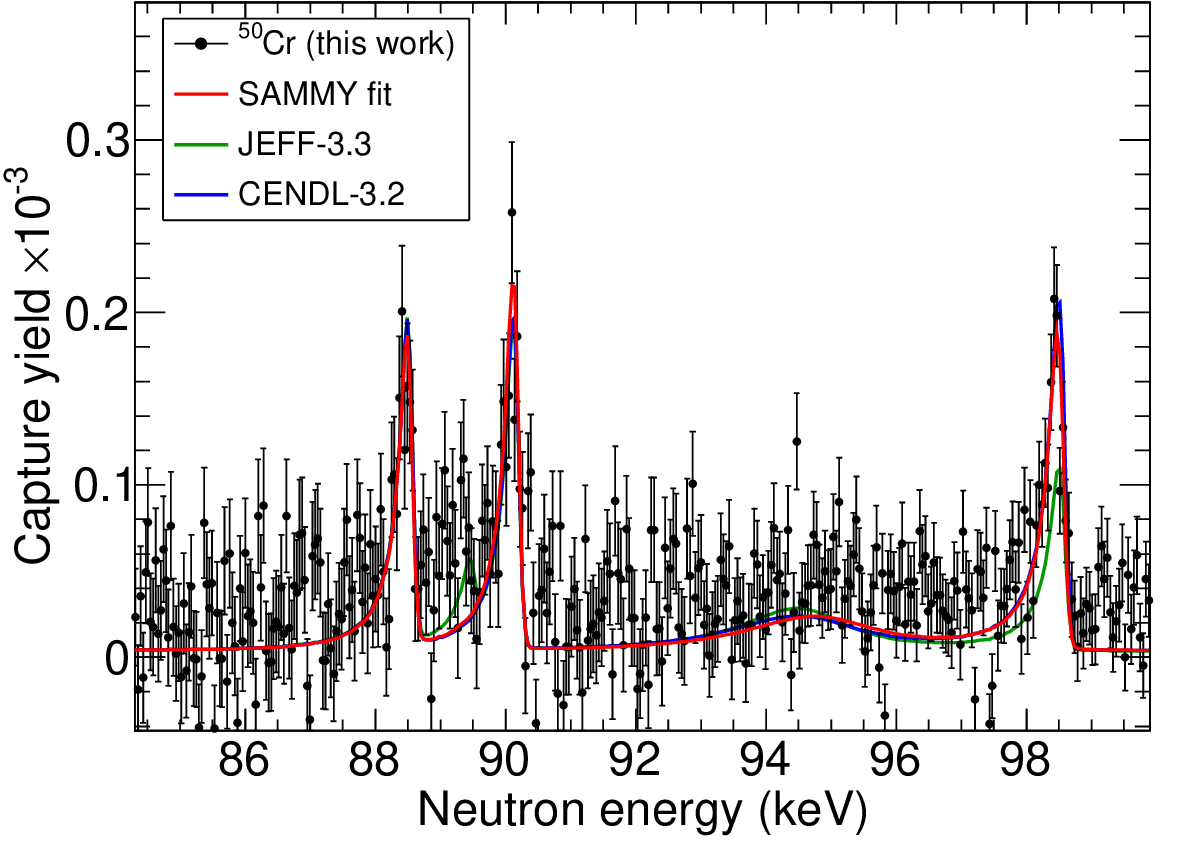}
      \includegraphics[width=0.5\linewidth]{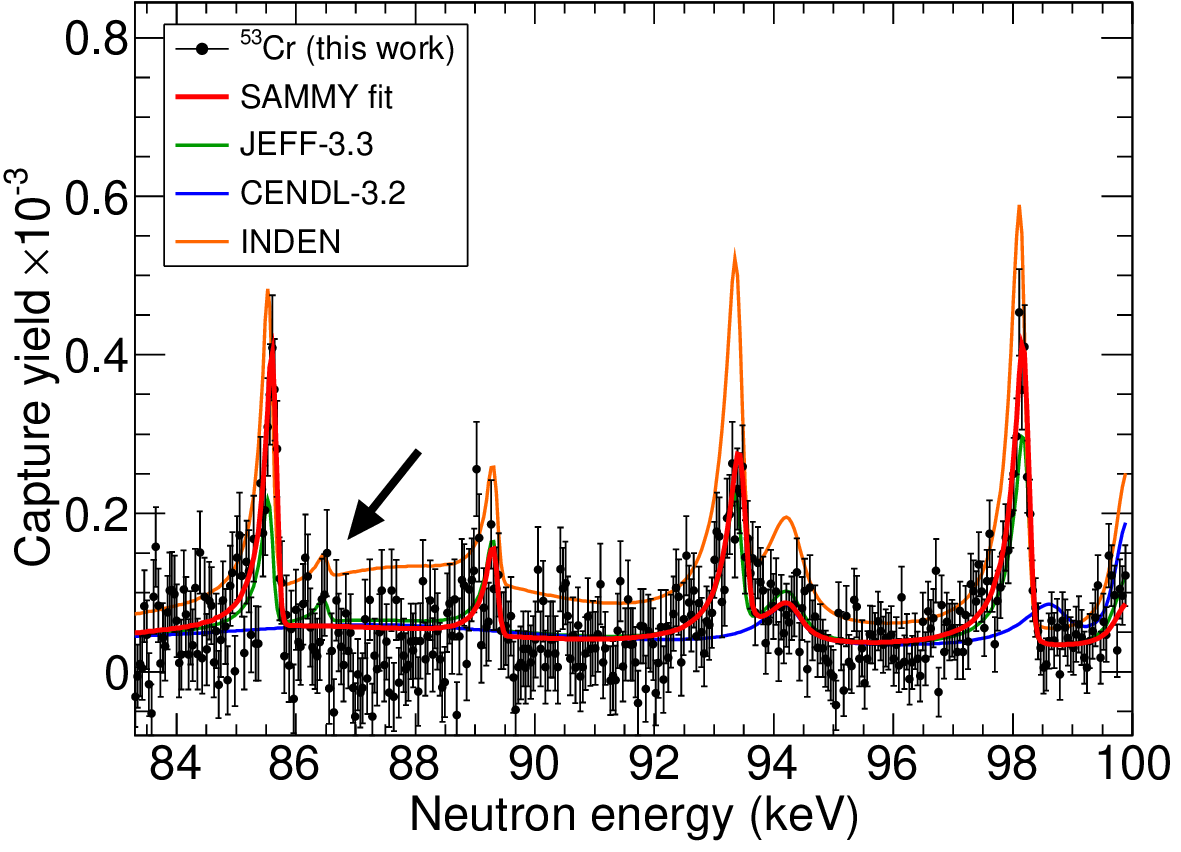}
      \caption{\small $^{50}$Cr (left) and $^{53}$Cr (right) capture yield fitted with SAMMY, along with the one predicted using JEFF-3.3, CENDL-3.2 and INDEN resonance parameters for comparison. The arrows indicate some resonance not present in the evaluations but visible in the n\_TOF data in the case of $^{50}$Cr, or the opposite situation in the case of $^{53}$Cr. See text for more details.} \label{plot-cr50-res}
\end{figure}

The results for $^{50}$Cr (see the left panels of Fig. \ref{plot-cr50-res}) evidence clear discrepancies between our data, well reproduced by the SAMMY fits, and the evaluations, of which JEFF-3.3 and INDEN share the same resonance parameters. Furthermore, we report 3 resonances of $^{50}$Cr, clearly visible in our data (see for example the one indicated with an arrow at 64~keV) but not present in JEFF-3.3. These were reported in JEFF-3.1~\cite{koning2006jeff}, but got removed from the newer versions of the library like JEFF-3.2~\cite{plompen2020joint} and JEFF-3.3.

In the case of $^{53}$Cr (see right panels of Fig. \ref{plot-cr50-res}), there are again clear differences between our data and the evaluations. The comparison indicates a clear overestimation of the cross section by the INDEN evaluation. Furthermore, there are 9 resonances of $^{53}$Cr (two of them marked with arrows at 41.8 and 86.2~keV) included in JEFF-3.3 and INDEN that cannot be observed in our data. As these resonances are very weak, we can neither confirm nor deny their existence due to low statistics. We have not included them in our list of resonance parameters. 

\subsection{Radiative kernels}
\label{subsect_kernel}
\begin{figure}
      \includegraphics[width=0.5\textwidth]{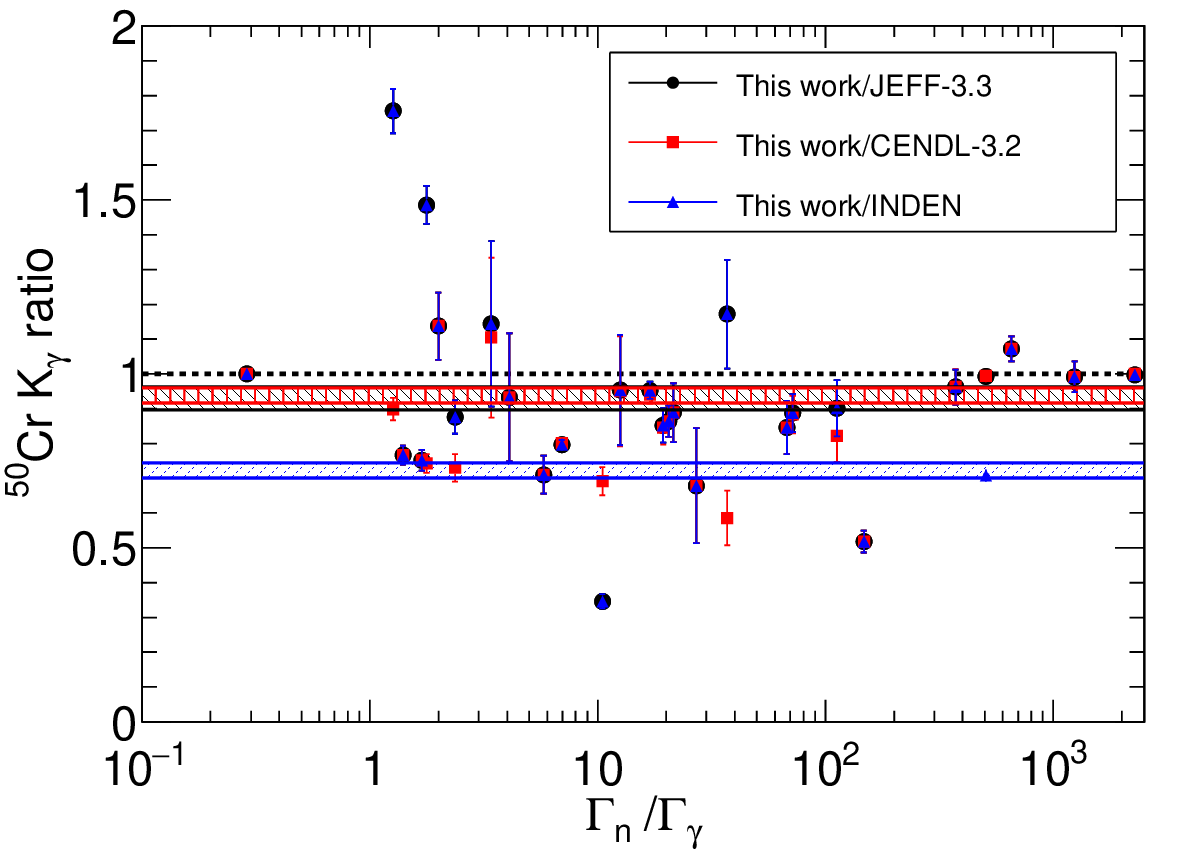}
      \includegraphics[width=0.5\textwidth]{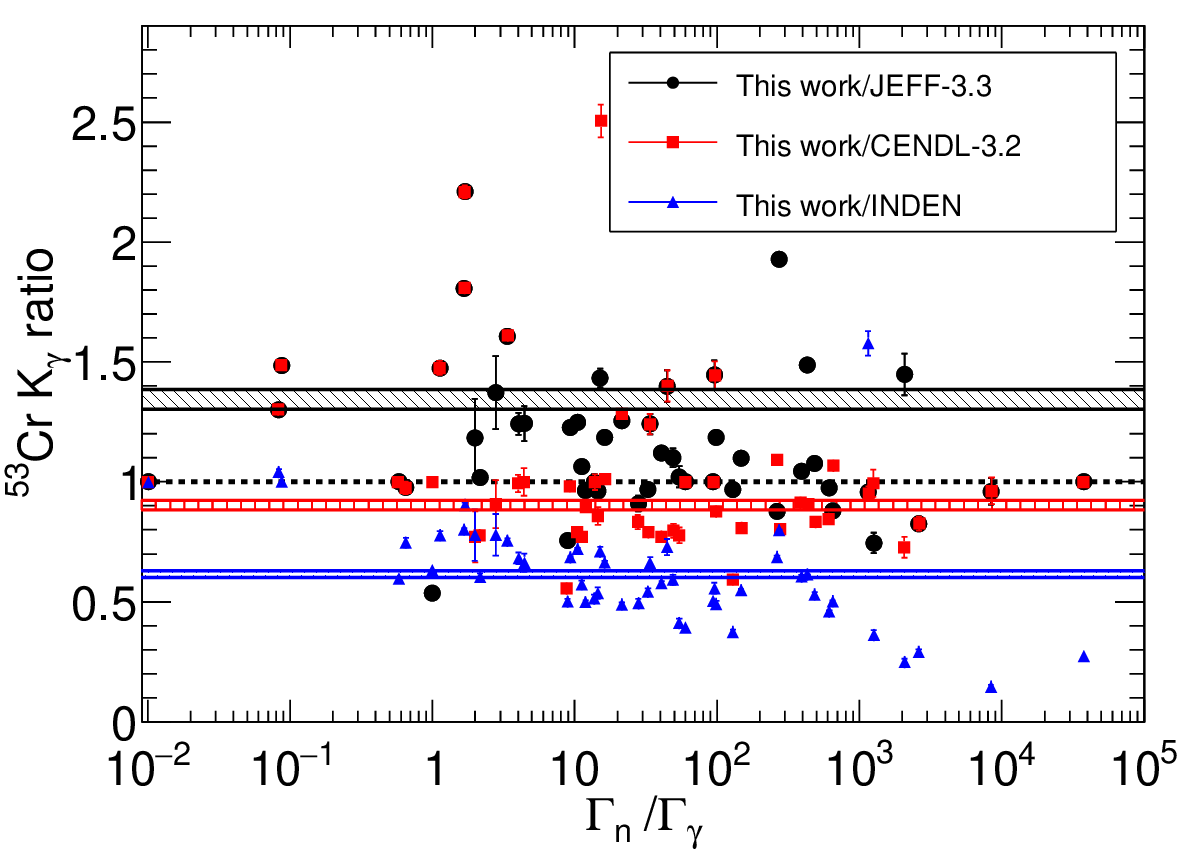}
      \caption{\small Ratio of radiative kernels obtained in this work and reported in the different evaluations as function of the scattering-to-capture ratio. The solid lines represent the standard deviation of the weighted mean value. See text for more details.} \label{plot-kernel-ratio}
\end{figure}

A quantitative comparison of the measured and evaluated cross sections in the RRR can be made  by using the radiative kernel $K_\gamma$, which is proportional to the integral of the resonance and is defined as
\begin{equation}
\label{eq-kernel}
K_\gamma = g_J \frac{\Gamma_\gamma\Gamma_n}{\Gamma_\gamma+\Gamma_n},
\end{equation}
where $g_J$ is the spin factor $g_J=\frac{2J+1}{(2i+1)(2I+1)}$, with $i=1/2$ and $I$ the spin of the neutron and the target nucleus respectively, and $J$ the total angular momentum of the resonance. We have compared the ratio between the $K_\gamma$ from this work and the evaluations as a function of the scattering-to-capture probability $\Gamma_n/\Gamma_\gamma$. This way, we can identify issues related to neutron scattering. The result is shown in Fig. \ref{plot-kernel-ratio} along with the weighted mean value of the ratio, using the uncertainty of the widths fitted by SAMMY as the weighting factor (only the resonances for which at least one width have been fitted are considered in the calculations).

The values for $^{50}$Cr show an overestimation of the previous evaluations with respect to this work, of about 7\% for JEFF-3.3 and CENDL-3.2 and almost 40\% for INDEN. The deviation is dominated by the main \textit{s}-wave resonance at 5.64~keV ($\Gamma_n/\Gamma_\gamma\approx500$), which is in perfect agreement with the first two libraries but 40\% lower than INDEN (as mention above, from 10~keV onwards JEFF-3.3 and INDEN share the same resonance parameters). We must point out that the \textit{p}-wave resonances at 5.46 and 9.31~keV have been omitted from this analysis because their kernel varies so much between evaluations that the ratios deviate about a factor 4 from the average, probably because their parameters are greatly influenced by the main \textit{s}-wave resonance. There is not any clear trend as a function of the scattering-to-capture ratio.

For the $^{53}$Cr case, where the evaluations feature important differences, our kernels are on average 35\% larger than JEFF-3.3, 10\% lower than CENDL-3.2 and around 60\% lower than INDEN. It is important to notice that our data deviates much more from the evaluations based on the data from Guber et al.~\cite{guber2011neutron} (JEFF-3.3 and INDEN) than from CENDL-3.2, based on the older measurements. In addition, there is an apparent decrease of the kernel ratio as a function of $\Gamma_n/\Gamma_\gamma$ for the INDEN case. This could indicate an overestimation of the neutron-scattering effects by this evaluation, but the lack of statistics in some of the resonances makes it difficult to completely confirm this issue, so a revision by the evaluators is recommended.

\subsection{Integral cross sections (MACS)}
\label{subsect_comp3}
As discussed in Ref.~\cite{perez2025neutron}, an integral cross section measurement is very useful to assess evaluations which present large discrepancies. In our case, the Maxwellian Averaged Cross Section at 30~keV ($\MACS$), commonly used in stellar nucleosynthesis calculations, is especially valuable because the chromium resonances contributing to benchmark experiments also strongly contribute to this MACS. We have calculated the $\MACS$ from our new cross section and the evaluations as
\begin{equation}
\label{eq-MACS}
\text{MACS}_{kT} = \frac{2}{\sqrt{\pi}}\frac{ \int_0^\infty E_ne^{-E_n/kT}\sigma(E_n)\text{d}E_n }{ \int_0^\infty E_ne^{-E_n/kT}\text{d}E_n },
\end{equation}
where $E_n$ is the neutron energy in the centre-of-mass system and $\sigma(E_n)$ is the point-wise capture cross section. Since the measurement described herein only covers the region below $E_n=100~$keV, we have used JEFF-3.3 and CENDL-3.2 to extrapolate the measured cross section up to 300~keV and evaluate the difference in the MACS when using one or another evaluation for the extrapolation. 

The results for $^{50}$Cr are 34.3 and 35.6~mb when extrapolating with JEFF-3.3 or CENDL-3.2, respectively. The same for $^{53}$Cr yields MACS values of 30.9 and 30.8~mb. In both cases the values are in agreement within a few percent and thus the average is adopted:
\begin{align}
\MACS(^{50}\text{Cr})=35.0(24)~\text{mb},\\
\MACS(^{53}\text{Cr})=30.9(22)~\text{mb}.
\end{align}
\label{eq-macs-tof}
The uncertainty is difficult to estimate, because it is not only related to that of the capture yield but also to the fitting process with SAMMY. Since the uncertainty of the yield varies between 5\% and 9\%, a value of 7\% has been adopted as the uncertainty of the MACS. 

\begin{figure}
      \includegraphics[width=0.5\textwidth]{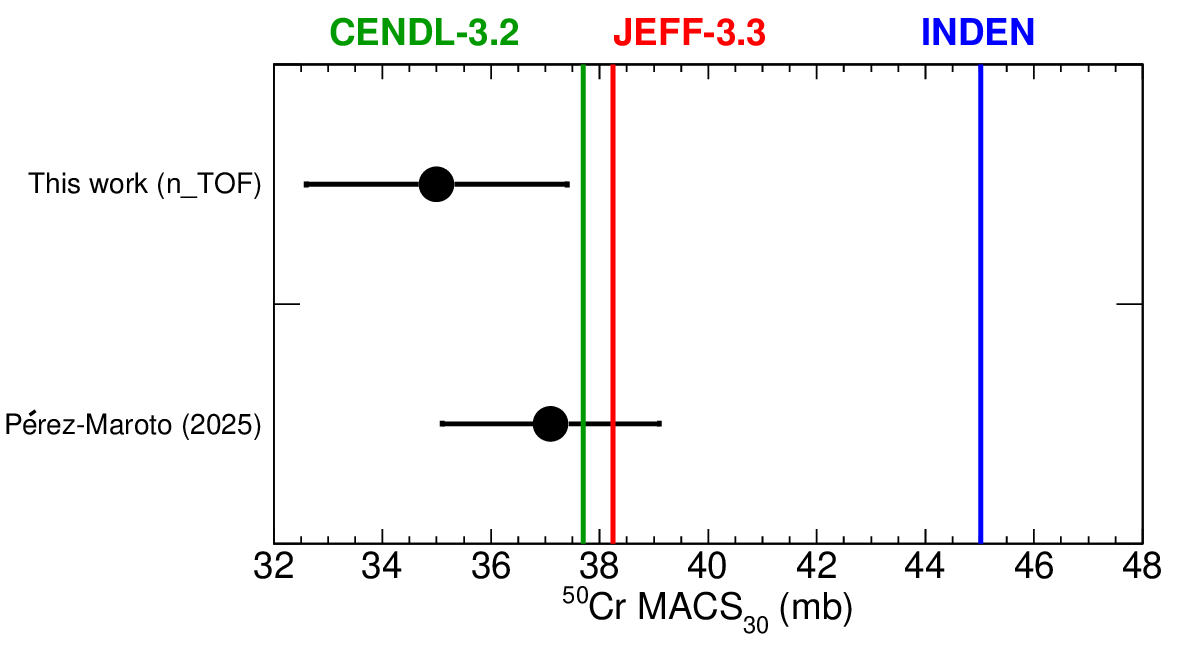}
      \includegraphics[width=0.5\textwidth]{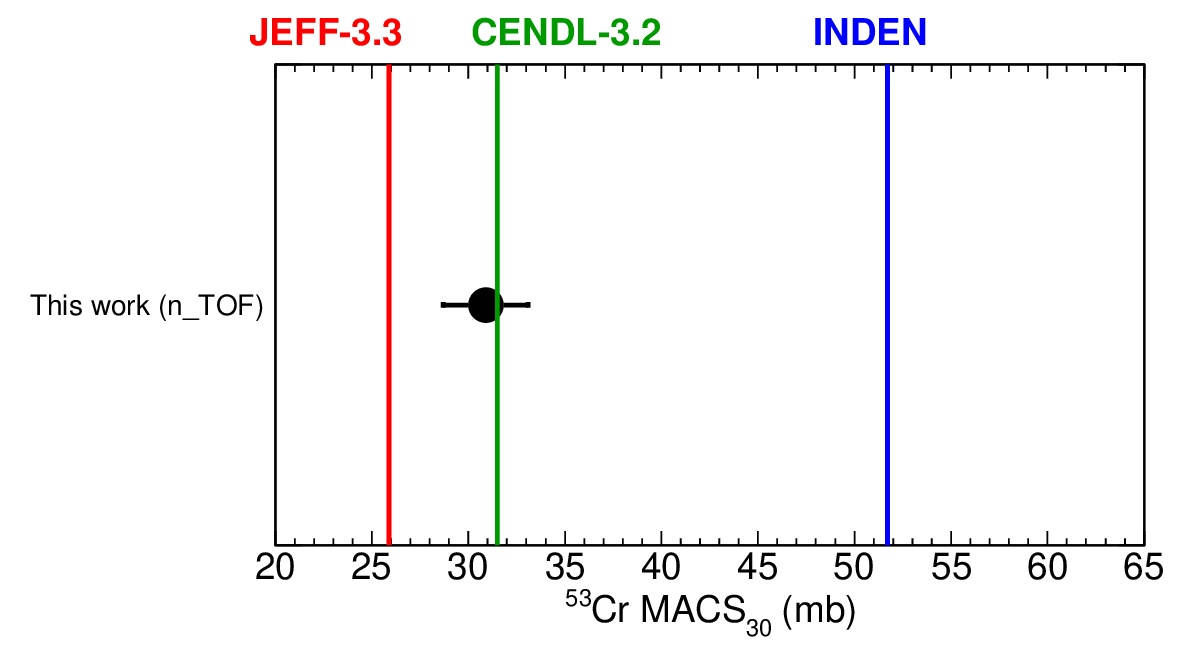}
      \caption{\small $\MACS$ comparison between the values obtained in this work for $^{50,53}$Cr, the $^{50}$Cr activation measurement and the values derived from evaluations.} \label{plot-macs-conc}
\end{figure}

Our results are compared in Table \ref{table-final-macs} and Fig. \ref{plot-macs-conc} to the MACS calculated from the cross sections libraries and from the only experimental value of the MACS of $^{50}$Cr by Pérez-Maroto et al.~\cite{perez2025neutron}. 
\begin{itemize}
    \item Regarding $^{50}$Cr, our $\MACS$ is in clear disagreement (30\% lower) with the recent INDEN evaluation, and 8-9\% lower, slightly beyond one standard deviation, than JEFF-3.3 and CENDL-3.2. Our value is then in a remarkable agreement within uncertainties with the $\MACS$ value of Pérez-Maroto et al. obtained recently by activation at the CNA HiSPANoS facility. 
    \item Regarding $^{53}$Cr, our data is only in agreement with the CENDL-3.2 evaluation. A significant disagreement is found with both JEFF-3.3 (20\% lower than our result) and INDEN (70\% larger) evaluations.
\end{itemize} 

These results confirm the trends that were observed when studying the radiative kernels, suggesting clearly that the recent increase in the neutron capture cross section proposed by INDEN is inappropriate for both $^{50}$Cr and $^{53}$Cr.

\begin{table}
\centering
\caption{\small $\MACS$ values of $^{50}$Cr and $^{53}$Cr obtained in this work, compared to the values extracted from the evaluations and the only $^{50}$Cr activation measurement available.}
\begin{tabular}{lcc}
\toprule
 MACS$_{30}$ (mb) &  $^{50}$Cr & $^{53}$Cr \\
 \midrule
     \underline{Evaluations}:\\
      JEFF-3.3 (2017)    & 38.2       & 25.9 \\
      CENDL-3.2 (2020)   & 37.7       & 31.5 \\ 
      INDEN (2023)       & 45.0       & 51.7 \\
      \underline{Activation @HiSPANoS}:\\
      Pérez-Maroto et al. (2025) & 37.1(20) & -  \\
      \underline{\textbf{\textit{ToF} @n\_TOF}}:\\
      \textbf{This work}   & \textbf{35.0(24)} & \textbf{30.9(22)}  \\
      \botrule
\end{tabular}
\label{table-final-macs}
\end{table}

\section{Summary and conclusions}
\label{sect_conc}
A neutron capture time-of-flight measurement of $^{50}$Cr and $^{53}$Cr has been successfully performed at the n\_TOF-EAR1 facility of CERN. The capture set-up consisted on four C$_6$D$_6$ Total Energy detectors characterized by their very low neutron sensitivity. Two highly enriched chromium oxide samples were used for each isotope: a very thin one was used to analyse the yield between 1 and 10~keV, and a thicker one for the rest of the neutron energy range up to 100~keV. All samples were thinner than any of those used in previous measurements, thus strongly suppressing the multiple-scattering effects that affected the previous measurements causing the discrepancies present in the evaluated cross section libraries.

The capture yield has been measured from 1 to 100~keV with an accuracy between 5 and 9\%, fulfilling the requirements of the NEA HPRL request. A total of 33 resonances of $^{50}$Cr and 51 resonances of $^{53}$Cr have been identified and analysed with SAMMY, resulting in a new set of resonance parameters. Overall, the comparison of radiative kernels indicates a $^{50}$Cr capture cross section $\sim$7\% lower than JEFF-3.3 (and thus ENDF/B-VIII.0, JENDL-5) and CENDL-3.2, and almost 40\% lower than INDEN (and thus ENDF/B-VIII.1). Very similar difference is also reached when the comparison is made for the $\MACS$ integral cross section. On the other hand, the $\MACS$ extracted from the $^{50}$Cr n\_TOF differential cross section measurement is in an excellent agreement with that obtained recently by neutron activation at CNA HiSPANoS~\cite{perez2025neutron}. 

The discrepancies with the evaluations are much larger in the case of $^{53}$Cr. According to the radiative kernels, our cross section is on average 35\% larger than JEFF-3.3 (and thus ENDF/B-VIII.0, JENDL-5), 10\% lower than CENDL-3.2 and a remarkable 60\% lower than INDEN (and thus ENDF/B-VIII.1). When comparing the $\MACS$ values, a very good agreement if found with CENDL-3.2, while our result is 70\% larger than the one expected from INDEN. 

The sizeable increase in the chromium cross sections proposed in the recent INDEN evaluation is not supported by the results shown herein. Therefore, a re-evaluation of these cross sections using the new experimental data might contribute to solve the actual issues with the criticality safety benchmarks sensitive to chromium and stainless steel.

\bmhead{Acknowledgements}
This measurement has received funding from the Euroatom research and training programme 2014-2018 under grant agreement No 847594 (ARIEL), and from the Spanish national projects RTI2018-098117-B-C21, PID2019-104714GB-C22, PID2021-123879OB-C21 and PID2022-138297NB-C21.

This work is also part of the PhD thesis of P. Pérez-Maroto at Universidad de Sevilla, Spain~\cite{PerezMaroto:2920339}, funded through the FPI national Grant No PRE2019-089678.

We also acknowledge the National Science Centre, Poland (Grant No. UMO-2021/41/B/ST2/00326). Support of funding agencies of all other institutes from the n\_TOF Collaboration is gratefully acknowledged.

\begin{appendices}

\section{Resonance parameters and kernels}
\label{appen}

\begin{table}[]
\centering
\caption{Resonance parameters of $^{50}$Cr obtained in this work. The uncertainty of the resonance energy has been considered as 0.1\% based on time-to-energy calibration. The uncertainty of the resonance widths is the one obtained in the SAMMY fitting. If a value has no uncertainty it means that it has not been fitted but adopted from an evaluated library (see Section \ref{subsect_comp2}).}
\begin{tabular}{ ccccccc }
\toprule
	   E (keV)  & J & $\ell$ & $\Gamma_{\gamma}$ (eV) & $\Gamma_n$ (eV) & Corr. (\%) & $K_\gamma$ (eV)\\ \midrule
      5.464(5) & 1.5 & 1 & 0.9(3) & 0.0315(6) & -20 & 0.61(3) \\
      5.581(6) & 0.5 & 0 & 3.226(15) & 1636(11) & 21 & 3.220(16)\\
      9.306(9) & 0.5 & 1 & 1.2(7) & 0.0539(18) & -28 & 0.0516(18)\\
      18.63(2) & 1.5 & 1 & 0.318(5) & 5.400 &- & 0.601(16)\\
      19.23(2) & 0.5 & 1 & 0.394(8) & 2.740 &- & 0.344(5)\\
      21.85(2)$^{\ast}$ & 0.5 & 1 & 0.370 & 0.039(7) &- & 0.035(5)\\ 
      24.08(2)$^{+@}$ & 1.5 & 1 & 0.170 & 0.049 & -& 0.076\\ 
      24.88(2)$^{\bullet}$ & 1.5 & 1 & 0.203(9) & 0.360 &- & 0.260(10)\\ 
      28.48(3) & 0.5 & 1 & 0.598(20) & 392 & -& 1.85(8)\\
      33.49(3) & 1.5 & 1 & 0.454(13) & 9.300 & -& 0.87(5) \\
      35.48(3) & 1.5 & 1 & 0.53(3) & 1.250 & -& 0.74(4)\\
      35.65(4) & 0.5 & 1 & 0.48(3) & 5.000 & -& 0.436(25)\\
      37.57(4) & 0.5 & 0 & 1.85(8) & 2300 &- & 0.75(4)\\
      40.65(4) & 1.5 & 1 & 0.348(16) & 39(7) &- & 0.69(6)\\
      46.74(5) & 1.5 & 1 & 0.53(3) & 0.900 &- & 0.67(3)\\
      50.07(5) & 1.5 & 1 & 0.271(14) & 1.570 &- & 0.462(8)\\
      53.62(5)$^{\bullet}$ & 1.5 & 1 & 0.59(6) & 1.180 &- & 0.79(7) \\ 
      55.13(5) & 0.5 & 0 & 0.76(4) & 281 &- & 0.75(4)\\
      59.21(6) & 1.5 & 1 & 0.578(19) & 11.200 &- & 0.110(6)\\
      63.29(6) & 1.5 & 1 & 0.270(19) & 10.000 &- & 0.53(7)\\
      64.08(6)$^{\ast\dagger}$ & 0.5 & 1 & 0.370 & 0.250 &- & 0.149\\ 
      64.85(6) & 0.5 & 0 & 0.60(4) & 43.000 &- & 0.59(4) \\
      65.87(7) & 1.5 & 1 & 0.490(23) & 33.100 &- & 0.96(9)\\
      68.24(7) & 0.5 & 1 & 0.44(13) & 1.800 &- & 0.35(7)\\
      70.28(7) & 0.5 & 1 & 0.88(9) & 1.230 &- & 0.513(18)\\
      73.35(7) & 1.5 & 1 & 0.442(23) & 9.500 &- & 0.84(8)\\
      75.38(7)$^{\ast}$ & 0.5 & 1 & 0.46(5) & 4.200 &- & 0.41(4)\\ 
      77.86(8) & 0.5 & 1 & 0.52(14) & 14.200 & -& 0.50(12)\\
      79.08(8) & 0.5 & 1 & 0.68(4) & 100 &- & 0.67(4)\\
      88.69(9) & 1.5 & 1 & 0.43(4) & 5.340 &- & 0.79(13)\\
      90.33(9) & 1.5 & 1 & 0.64(11) & 2.170 &- & 0.98(20)\\
      94.95(9)$^{@}$ & 0.5 & 0 & 0.967 & 2200 &- & 0.967 \\ 
      98.71(10) & 0.5 & 1 & 1.82(21) & 2.300 &- & 1.02(4)\\ \botrule
      \multicolumn{5}{l}{\footnotesize{$+)$ $\Gamma_{\gamma}$ and $\Gamma_n$ from JEFF-3.3}} \\
      \multicolumn{5}{l}{\footnotesize{$@)$ $\Gamma_{\gamma}$ and $\Gamma_n$ from CENDL-3.2}} \\
      \multicolumn{5}{l}{\footnotesize{$\dagger)$ $\Gamma_{\gamma}$ and $\Gamma_n$ from JEFF-3.1}} \\      
      \multicolumn{5}{l}{\footnotesize{$\bullet)$ Energy discrepancy}} \\
      \multicolumn{5}{l}{\footnotesize{$\ast)$ Removed from JEFF-3.2}} \\
\end{tabular}
\label{table-cr50-res}
\end{table}

\begin{table}[]
\centering
\caption{Resonance parameters of $^{53}$Cr obtained in this work. The uncertainty of the resonance energy has been considered as 0.1\% based on time-to-energy calibration. The uncertainty of the resonance widths is the one obtained in the SAMMY fitting. If a value has no uncertainty it means that it has not been fitted but adopted from an evaluated library (see Section \ref{subsect_comp2}).}
\begin{tabular}{ ccccccc }
\toprule
	   E (keV)  & J & $\ell$ & $\Gamma_{\gamma}$ (eV) & $\Gamma_n$ (eV) & Corr. (\%) & $K_\gamma$ (eV)\\ \midrule
      4.033(4) & 1 & 0 & 3.089(12) & 1332(8) & 7 & 1.1558(17)\\
      5.677(7) & 2 & 0 & 0.541(7) & 143(3) & 29 & 0.337(3)\\
      6.857(7) & 1 & 0 & 3.31(3) & 906(13) & 53 & 1.236(5)\\
      8.210(8) & 2 & 0 & 1.680(20) & 1091(18) & 35 & 1.049(8)\\
      12.11(1) & 2 & 1 & 0.430(6) & 4.500 &- & 0.2452(17)\\
      12.96(1) & 2 & 1 & 0.212(7) & 0.59(6) & -64 & 0.097(11)\\
      14.66(1)$^{\bullet}$ & 1 & 1 & 0.343(7) & 3.200 &- & 0.1161(7)\\ 
      16.04(2)$^{\ast}$ & 1 & 1 & 0.029(3) & 0.393 &- & 0.0100(3)\\ 
      19.71(2) & 2 & 0 & 0.708(11) & 104 &- & 0.439(4)\\
      20.23(2) & 3 & 1 & 0.406(6) & 6.600 &- & 0.335(4)\\
      22.49(2) & 2 & 1 & 0.208(19) & 0.840 &- & 0.104(4)\\
      24.34(2)$^{\ast}$ & 0 & 1 & 0.58(5) & 0.66(6) & -44 & 0.0388(8)\\ 
      25.90(3) & 2 & 0 & 0.40(3) & 195 & -& 0.250(5)\\
      27.28(3) & 1 & 0 & 1.11(3) & 680 & -& 0.416(4)\\
      28.79(3)$^{\ast}$ & 1 & 1 & 5.7(6) & 0.478(15) &-13& 0.1656(19)\\ 
      28.89(3) & 0 & 1 & 1.85(10) & 4.0(4) & -64 & 0.158(4)\\
      29.49(3) & 2 & 0 & 0.894(19) & 350 &- & 0.557(7)\\
      31.60(3) & 3 & 1 & 0.268(8) & 8.748 &- & 0.227(5)\\
      32.17(3) & 3 & 1 & 0.241(8) & 2.700 &- & 0.194(5)\\
      35.04(3) & 3 & 1 & 0.221(12) & 3.210 &- & 0.181(8)\\
      37.84(4) & 3 & 1 & 0.245(11) & 12.000 & -& 0.210(8)\\
      42.47(4) & 1 & 1 & 0.521(25) & 6.210 &- & 0.180(3)\\
      43.30(4) & 3 & 1 & 0.34(3) & 1.500 &- & 0.242(14)\\
      47.22(5) & 1 & 1 & 0.86(4) & 34.900 & -& 0.313(5)\\
      47.79(5)$^{\ast+}$ & 3 & 1 & 2.427 & 0.025 & -& 0.021 \\ 
      49.97(5) & 3 & 1 & 0.257(11) & 7.200 & -& 0.217(8)\\
      51.07(5) & 1 & 1 & 0.46(3) & 45.100 &- & 0.169(4)\\
      53.63(5) & 1 & 1 & 0.93(4) & 20.000 &- & 0.334(5)\\
      55.27(5)$^{+}$ & 0 & 1 & 0.520 & 31.127 &- & 0.064\\ 
      55.87(6)$^{+}$ & 0 & 1 & 0.520 & 48.510 &- & 0.064\\ 
      62.39(6)$^{@}$ & 2 & 1 & 0.300 & 0.300 &- & 0.097\\ 
      62.63(6) & 1 & 1 & 1.43(7) & 4.817 &- & 0.414(5)\\
      64.83(6) & 2 & 1 & 0.53(3) & 4.657 &- & 0.294(8)\\
      65.51(7) & 2 & 0 & 0.48(4) & 4017 &- & 0.300(17)\\
      66.50(7) & 3 & 1 & 0.308(17) & 13.698 &- & 0.263(12)\\
      67.15(7) & 2 & 1 & 0.44(3) & 15.029 &- & 0.269(9)\\
      69.51(7) & 1 & 1 & 0.92(6) & 50.100 &- & 0.340(14)\\
      69.81(7) & 2 & 1 & 1.86(11) & 3.112 & -& 0.728(6)\\
      73.01(7) & 1 & 1 & 0.59(4) & 76.000 &- & 0.219(6)\\
      73.60(7) & 1 & 0 & 1.04(9) & 1200 & -& 0.391(12)\\
      74.43(7) & 3 & 1 & 0.89(6) & 1.76(18) & -49 & 0.52(7)\\
      74.53(7) & 2 & 0 & 0.48(5) & 1000 &- & 0.301(18)\\
      75.67(8)$^{\dagger}$ & 2 & 1 & 2.102 & 0.184 &- & 0.106\\ 
      76.62(8)$^{+}$ & 1 & 1 & 0.494 & 0.287 &- & 0.681\\ 
      85.80(9) & 1 & 1 & 0.193(16) & 3.272 &- & 0.455(6)\\
      86.99(9) & 1 & 0 & 2.48(24) & 6500 &- & 0.93(3)\\
      89.50(9) & 2 & 1 & 0.65(6) & 0.42(4) & -5 & 0.160(4)\\
      93.64(9) & 2 & 0 & 0.82(5) & 78.533 &- & 0.507(21)\\
      94.56(9) & 2 & 0 & 0.40(4) & 501 &- & 0.248(14)\\
      95.98(10)$^{+@}$ & 0 & 1 & 0.890 & 33446 &- & 0.111\\
      98.40(10) & 2 & 1 & 1.30(6) & 19.570 &- & 0.759(21)\\ \botrule
      \multicolumn{5}{l}{\footnotesize{$+)$ $\Gamma_{\gamma}$ and $\Gamma_n$ from JEFF-3.3}} \\
      \multicolumn{5}{l}{\footnotesize{$@)$ $\Gamma_{\gamma}$ and $\Gamma_n$ from CENDL-3.2}} \\
      \multicolumn{5}{l}{\footnotesize{$\dagger)$ $\Gamma_{\gamma}$ and $\Gamma_n$ from INDEN}} \\
      \multicolumn{5}{l}{\footnotesize{$\bullet)$ Energy discrepancy}} \\
      \multicolumn{5}{l}{\footnotesize{$\ast)$ Not included in CENDL-3.2}} \\
\end{tabular}
\label{table-cr53-res2}
\end{table}




\end{appendices}


\bibliography{sn-bibliography}

\end{document}